\documentclass[12pt,letterpaper]{article}
\pdfoutput=1

\usepackage{graphicx,array}
\usepackage{color}
\usepackage{latexsym}
\usepackage{amsthm}
\usepackage{amsmath}
\usepackage{amssymb}
\usepackage[numbers,sort&compress]{natbib}
\usepackage{bm}
\usepackage{slashed} 
\usepackage{mathrsfs}
\usepackage{lipsum}
\usepackage{hyperref} 
\hypersetup{
    colorlinks=true,       
    linkcolor=red,          
    citecolor=blue,        
    filecolor=magenta,      
    urlcolor=blue           
}
\usepackage[all]{hypcap} 

\usepackage{myterms}

\newcommand{\ctime}{\eta}
\newcommand{\metric}{\textit{g}}
\newcommand{\eff}{\mathrm{eff}}
\newcommand{\cmb}{\mathrm{cmb}}
\newcommand{\RH}{\text{\sc rh}}
\newcommand{\drho}{\delta \hspace{-0.03cm} \rho}

\title{\bf Superheavy scalar dark matter \\ from gravitational particle production \\ in $\alpha$-attractor models of inflation}

\author{\large Siyang Ling and Andrew J. Long}
\date
{\small \it 
Department of Physics and Astronomy, Rice University, Houston, Texas 77005, U.S.A.
}

\begin{document}

\maketitle

\begin{abstract}
We study the phenomenon of gravitational particle production as applied to a scalar spectator field in the context of $\alpha$-attractor inflation.  
Assuming that the scalar has a minimal coupling to gravity, we calculate the abundance of gravitationally-produced particles as a function of the spectator's mass $m_\chi$ and the inflaton's $\alpha$ parameter.  
If the spectator is stable and sufficiently weakly coupled, such that it does not thermalize after reheating, then a population of spin-0 particles is predicted to survive in the universe today, providing a candidate for dark matter.  
Inhomogeneities in the spatial distribution of dark matter correspond to an isocurvature component, which can be probed by measurements of the cosmic microwave background anisotropies.  
We calculate the dark matter-photon isocurvature power spectrum and by comparing with upper limits from \textit{Planck}, we infer constraints on $m_\chi$ and $\alpha$.  
If the scalar spectator makes up all of the dark matter today, then for $\alpha = 10$ and $T_\RH = 10^4 \GeV$ we obtain $m_\chi > 1.8 \times 10^{13} \GeV \approx 1.2 \, m_\phi$, where $m_\phi$ is the inflaton's mass.

\end{abstract}

\newpage
\begingroup
\hypersetup{linkcolor=black}
\tableofcontents
\endgroup

\setlength{\parindent}{25pt}
\setlength{\parskip}{1.2ex}

\section{Introduction}\label{sec:Introduction}

Modern cosmology is characterized by an abundance of precision data.  
Cosmological surveys, including measurements of the cosmic microwave background and large scale structure, tell us how the universe evolved from the big bang and what makes up the universe today.  
These empirical observations indicate that most of the matter in the universe takes the form of an as-yet unidentified dark matter~\cite{Aghanim:2018eyx}, and measurements of inhomogeneities on cosmological scales indicate that the very early universe experienced a period of accelerated expansion known as cosmological inflation~\cite{Akrami:2018odb}.  
But despite the variety and precision of these cosmological observations, we are still left wondering about the nature of dark matter and the physics that drove inflation.  

There is no shortage of particle physics theories when it comes to candidates for dark matter~\cite{Bertone:2018krk} or models of inflation~\cite{Martin:2013tda}.  
If we allow ourselves to be guided through the vast landscape of theories by a principle of ``simplicity,'' then we would naturally be drawn to a scenario in which both the inflaton and the dark matter are scalar fields that only interact with one another and with the Standard Model particles through gravity.  
The inflaton potential would be simply $V(\phi) = m_\phi^2 \phi^2 / 2$ as in quadratic chaotic inflation~\cite{Linde:1983gd,Linde:1984st,Madsen:1988xe} and the dark matter would be generated via gravitational particle production~\cite{Chung:1998zb,Chung:1998ua,Kuzmin:1998kk}.  
In fact there have been many studies of superheavy (WIMPzilla) dark matter production in the context of quadratic inflation.  

In this work, we suppose that the early period of cosmological inflation can be described by the $\alpha$-attractor class of models~\cite{Kallosh:2013pby,Kallosh:2013lkr,Kallosh:2013hoa,Ferrara:2013rsa,Kallosh:2013maa,Kallosh:2013yoa,Galante:2014ifa,Roest:2015qya,Linde:2015uga}, and we suppose that dark matter is a spin-0 (scalar) particle that only interacts gravitationally, being produced through the phenomenon of gravitational particle production during inflation.  
We are interested in $\alpha$-attractors for several reasons.  
First off, in the last several years inflation with a quadratic potential has been excluded by the \textit{Planck} satellite's observations of the cosmic microwave background~\cite{Akrami:2018odb}.  
Alpha attractor models interpolate between the now-excluded quadratic inflation and a family of models that remain viable and testable with next-generation cosmological surveys.  
In particular, the scalar spectral index $n_s$ and the tensor-to-scalar ratio $r$ are predicted to obey the simple relations $n_s \approx 1 - 2/N_\cmb$ and $r \approx 12 \alpha / N_\cmb^2$~\cite{Kallosh:2013yoa}, which are consistent with both the measurement of $n_s$ for $N_\cmb \approx 50-60$ and the upper limit on $r$ for $\alpha < O(10)$.  
Moreover, certain value of $\alpha$ correspond to well-studied theories of inflation including Starobinsky inflation~\cite{Starobinsky:1980te} at $\alpha = 1$.  
In this sense, the $\alpha$-attractor provides a convenient parametrization for studying how the dark matter observables, \textit{i.e.} relic abundance and isocurvature, depend on the model of inflation.  
Moreover, $\alpha$-attractors models are interesting in their own right, and they arise in theoretically compelling theories of supergravity with a modified K\"ahler potential.  
Finally, the subject of gravitational dark matter production during $\alpha$-attractor inflation has not been studied before.  
Similar works in this line direction include nonthermal dark matter produced during reheating or preheating~\cite{Mishra:2017ehw,Allahverdi:2018iod,Maity:2018dgy,Garcia:2020eof}, gravitational-mediated inflaton decay \cite{Watanabe:2015eia}, gravitational production of gravitino dark matter in Starobinsky inflation~\cite{Addazi:2016bus,Addazi:2017kbx}, or gravitational dark matter production in Palatini preheating~\cite{Karam:2020rpa}.  

Using standard techniques from the study of quantum field theory in curved spacetime, we calculate the evolution of the inflaton field and the spacetime geometry during inflation and the period afterward, during which reheating occurs.  
Assuming a scalar spectator field that's non-conformally coupled to gravity, we calculate the spectrum, relic abundance today, and isocurvature of the spin-0 spectator field particles that are produced via the phenomenon of gravitational particle production.  

The outline of this article is as follows.  
We begin in \sref{sec:inflation} with a brief review of the $\alpha$-attractor models of inflation.  
Then in \sref{sec:GPP} we discuss inflationary gravitational particle production in the context of a minimally-coupled scalar field, which is a spectator field during inflation.  
Our main results appear in \sref{sec:DM}, where we study the evolution of the spectator field's mode functions in the inflationary spacetime background, we calculate the spectrum of gravitationally-produced particles, we infer their relic abundance today as a function of the reheating temperature, we calculate the dark matter isocurvature power spectrum, and we impose constraints on the parameter space of these theories.  
Finally \sref{sec:Conclusion} contains our summary and conclusion.  

\textit{Conventions.}  
We use natural units in which the speed of light and the reduced Planck constant are set equal to unity, $c = \hbar = 1$.  
We denote the reduced Planck mass by $M_p = 1 / \sqrt{8 \pi G_N} \simeq 2.435 \times 10^{18} \GeV$ where $G_N$ is Newton's constant.  
Our sign conventions for gravitational tensors correspond to $(-,+,+)$ in the Misner, Thorne, \& Wheeler scheme~\cite{MTW:1973}.  
In particular, the Minkowski spacetime metric is $\eta_{\mu\nu} \ud x^\mu \ud x^\nu = (\ud x^0)^2 - (\ud x^1)^2 - (\ud x^2)^2 - (\ud x^3)^2$.  

\section{$\alpha$-attractor models of inflation}\label{sec:inflation}

This section is a brief overview of $\alpha$-attractor models.  
We specify the inflaton potentials that define these models and discuss the viable parameter space.  
We numerically solve the inflaton's equation of motion for a fiducial parameter choice to illustrate the evolution of the inflaton field and the scale factor.  
The non-adiabatic evolution of the scale factor results in the phenomenon of gravitational particle production, which is discussed in the following section.  

\begin{figure}[t]
\begin{center}
\includegraphics[width=0.49\textwidth]{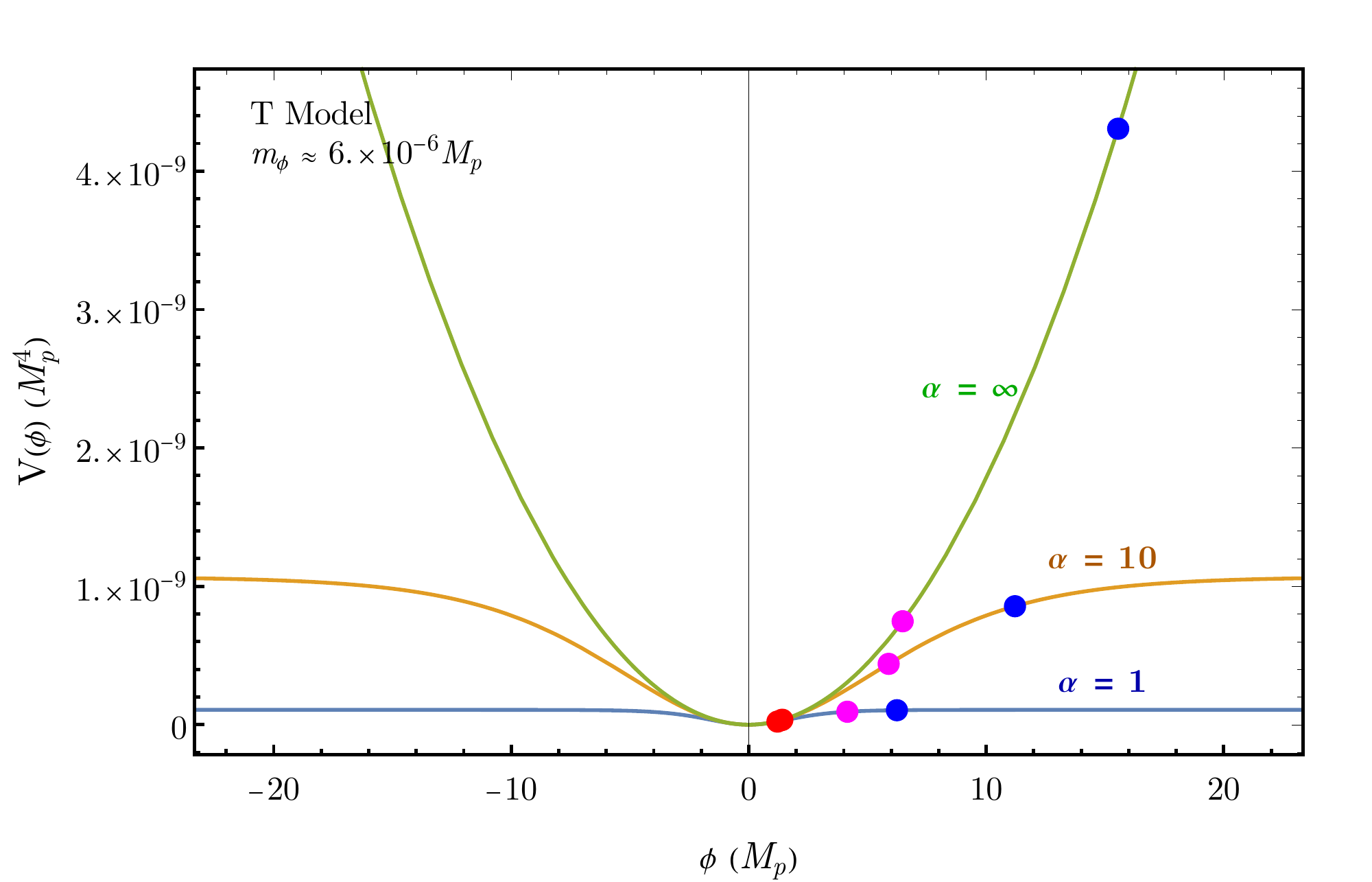} 
\includegraphics[width=0.49\textwidth]{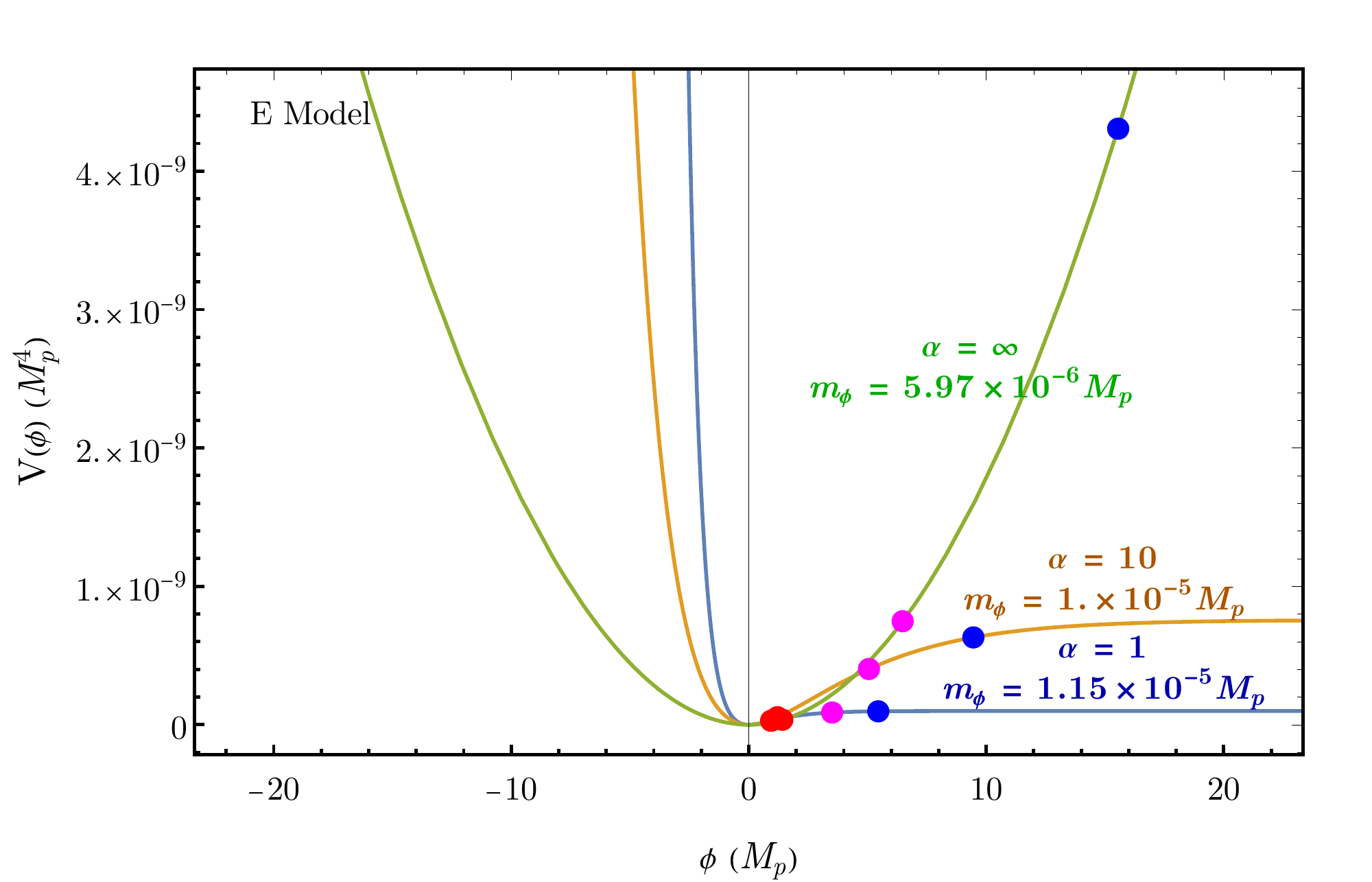} 
\caption{\label{fig:potential}\small
Inflaton potentials for the T-model and E-model $\alpha$-attractors.  Each panel shows several values of $\alpha$, and the value of $m_\phi$ is chosen such that $A_s$ agrees with the value measured by \textit{Planck}.  There are three dots along each curve; the red dot indicates the value of $\phi$ at the end of inflation, the magenta dot indicates $10$ $e$-foldings before the end of inflation, and the blue dot indicates $60$ $e$-foldings before the end of inflation.  
}
\end{center}
\end{figure}

We consider both the T-model and E-model $\alpha$-attractors.  
These models are specified by the scalar potentials~\cite{Kallosh:2013yoa,Ferrara:2013rsa,Kallosh:2015lwa} 
\begin{align}\label{eq:VT_and_VE}
	V_T(\phi) = \alpha \mu^2 M_p^2 \tanh^2 \frac{\phi}{\sqrt{6\alpha} M_p} 
	\qquad \text{and} \qquad 
	V_E(\phi) = \alpha \mu^2 M_p^2 \Bigl( 1 - e^{- \sqrt{2/3\alpha} \, \phi / M_p} \Bigr)^2 
	\com
\end{align}
where the inflaton field $\phi$ has a canonically-normalized kinetic term.  
Each model has two parameters:  a mass scale $\mu$ and a dimensionless parameter $\alpha$ that affects both the scale and shape of the inflaton potential.  
The inflaton's mass, $m_\phi \equiv \sqrt{V^\pprime(0)}$, is given by 
\begin{align}\label{eq:m_phi}
	m_\phi = \begin{cases}
	\mu / \sqrt{3} & , \quad \text{T-model} \\
	2 \mu / \sqrt{3} & , \quad \text{E-model} 
	\end{cases}
	\per
\end{align}
In \fref{fig:potential} we show the inflaton potential for several values of $\alpha$.  
Note how the $\alpha \to \infty$ limit reduces to a quadratic potential, $V_T, V_E \to m_\phi^2 \phi^2 / 2$.  
The $\alpha \ll 1$ regime gives a potential that flattens into a plateau where $\phi > O(\sqrt{\alpha} M_p)$.  

The inflaton's energy drives the cosmological expansion during inflation and reheating.  
Since the inflaton field is nearly homogeneous, we model the background geometry with the homogeneous and isotropic Friedmann-Robertson-Walker (FRW) spacetime.  
The metric is 
\begin{align}\label{eq:FRW}
	\ud s^2 
	= a(\ctime)^2 \, \bigl( \ud \ctime^2 - |\ud \xvec|^2 \bigr)
\end{align} 
where $a(\ctime)$ is the dimensionless scale factor, $\ctime$ is the comoving time coordinate (conformal time), and $\xvec$ is the comoving spatial coordinate 3-vector.  
Time derivatives of $a(\ctime)$ give the Hubble parameter $H(\ctime) = a^\prime / a^2$ and the Ricci scalar $R(\ctime) = - 6 a^\pprime / a^3$.  

\begin{figure}[t!]
\begin{center}
\includegraphics[width=\textwidth]{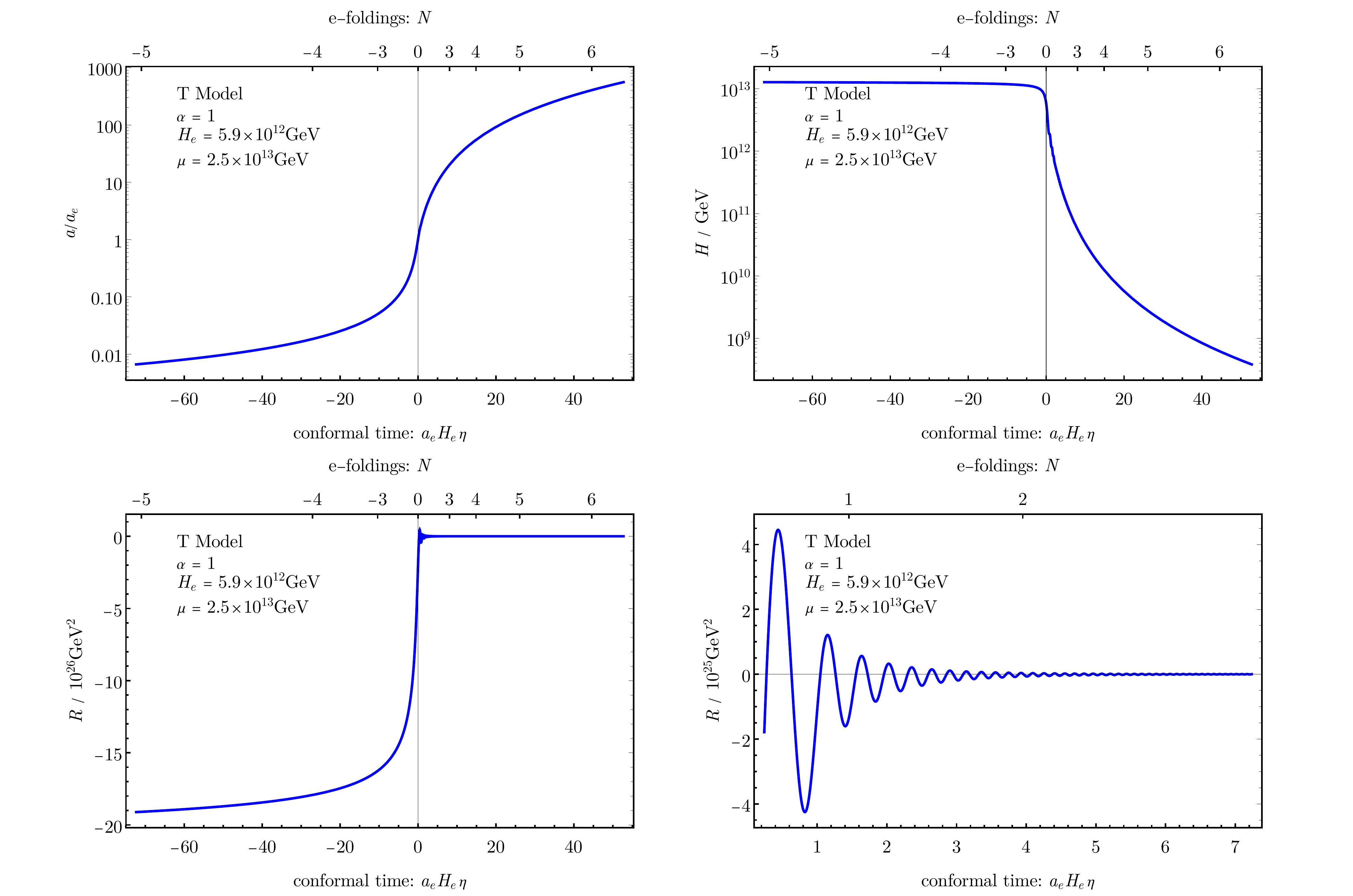} 
\caption{\label{fig:background}\small
Evolution of the spacetime background for a fiducial model of inflation.  We show the T-model $\alpha$-attractor with $\alpha = 1$ and $\mu = 2.5 \times 10^{13} \GeV$, implying $m_\phi = 1.5 \times 10^{13} \GeV$, $H_\cmb = 1.4 \times 10^{13} \GeV$, $H_e = 5.9 \times 10^{12} \GeV$, and $N_\cmb = 60$. The top panels show the FRW scale factor $a$ (top-left) and the Hubble parameter $H$ (top-right).  Both bottom panels show the Ricci scalar $R$ with the bottom-right panel showing a ``zoom'' to highlight oscillations near the end of inflation.  A vertical black line indicates the end of inflation at $\ctime = \ctime_e = 0$, and the top of the frame shows the number of e-foldings after the end of inflation $N(\ctime) = \ln a(\ctime) / a_e$.  
}
\end{center}
\end{figure}

The inflaton field equation and the Einstein field equation, when restricted to homogeneous field configurations and the FRW metric, lead to the following equations of motion:
\begin{equation}\label{eq:EOM}
	\phi^\pprime + 2 aH \phi^\prime + a^2 \frac{dV}{d\phi} = 0 
	\qquad \text{and} \qquad 
	3 M_p^2 H^2 = V + (\phi^\prime)^2 / (2a^2)
	\per
\end{equation}
Here $V(\phi)$ stands for either $V_T$ or $V_E$ from \eref{eq:VT_and_VE}.  
We show the FRW spacetime evolution in \fref{fig:background}.  
Note how $H(\ctime)$ and $R(\ctime)$ are approximately constant during inflation while $a(\ctime)$ grows.  
We define the fiducial ``end of inflation'' as the time $\ctime_\mathrm{end} \equiv \ctime_e$ when $d(1/aH)/d\ctime = 0$, such that the comoving Hubble scale $1/(aH)$ stops decreasing and begins increasing at the end of inflation.  
Without loss of generality, we shift the time coordinate such that $\ctime_e = 0$.  
After inflation has ended, $H$ and $R$ tend to decrease like a power law, but they also display an oscillatory component, which follows from the oscillations of the inflaton field around the minimum of its potential.  
In particular, note that $R$ oscillates to both positive and negative values.  
In the next section, we will discuss how the evolution of $a(\ctime)$ leads to the phenomenon of gravitational particle production in fields that are coupled to gravity non-conformally.  

The parameters $\alpha$ and $\mu$ must be chosen so that the $\alpha$-attractor model predicts a scalar power spectrum that's in agreement with the observed cosmic microwave background (CMB) temperature anisotropies.  
The comoving wavenumber of the CMB-scale modes is taken to be the pivot scale $k_\cmb = 0.002 \Mpc^{-1} a_0$ where $a_0 = a(\ctime_0)$ is the value of the scale factor today.  
The quantity of interest is the dimensionless ratio $k_\cmb / a_e H_e$ where $a_e \equiv a(\ctime_e)$ and $H_e \equiv H(\ctime_e)$.  
For the $\alpha$-attractor models that we consider, the universe is matter-dominated from the end of inflation until reheating at time $\ctime_\RH$, which implies~\cite{Liddle:2003as,Cook:2015vqa} 
\begin{align}\label{eq:a0_over_ae}
	\frac{a_0}{a_e} = \left( \frac{90 M_p^2 H_e^2}{\pi^2 g_{\ast S,0} T_0^3 T_\RH} \right)^{1/3} 
\end{align}
where $T_\RH$ is the plasma temperature at reheating, $T_0 \simeq 0.234 \meV$ is the temperature today, and $g_{\ast S,0} \simeq 3.91$.  
Combining these expressions gives 
\begin{align}\label{eq:k_cmb}
	\frac{k_\cmb}{a_e H_e} \simeq \bigl( 6.09 \times 10^{-24} \bigr) \left( \frac{H_e}{10^{13} \GeV} \right)^{-1/3} \, \left( \frac{T_\RH}{10^9 \GeV} \right)^{-1/3} 
\end{align}
where the numerical prefactor is approximately $e^{-53}$.  
The CMB-scale modes left the horizon at a time $\ctime_\cmb$ such that $k_\cmb = a_\cmb H_\cmb$ with $a_\cmb \equiv a(\ctime_\cmb)$ and $H_\cmb \equiv H(\ctime_\cmb)$.  
Solving this relation for $\ctime_\cmb$ yields the number of e-foldings between CMB horizon-crossing and the end of inflation, $N_\cmb = \ln (a_e / a_\cmb)$.  
For $H_e = 10^{13} \GeV$ and $T_\RH = 10^9 \GeV$ we have $N_\cmb \approx 54$.  

The predicted amplitude of the scalar power spectrum follows from a standard calculation~\cite{Baumann:2009ds}, and we find 
\begin{align}\label{eq:As}
	A_s & = \frac{\Acal}{8\pi^2} \frac{\alpha^2 \mu ^2}{M_p^2} \sinh^4 \left(\frac{\phi_\cmb}{\sqrt{6\alpha}M_p}\right) 
\end{align}
where $\Acal = 1$ for the T-model and $\Acal = 4$ for the E-model.  
The field amplitude at the time when the CMB-scale modes left the horizon is denoted by $\phi_\cmb \equiv \phi(\ctime_\cmb)$; it is given by
\begin{subequations}\label{eq:phi_N_cmb}
\begin{align}
	\phi_\cmb = \sqrt{\frac{3\alpha}{2}} \, M_p \ \mathrm{arcsech}\left(\frac{3 \alpha }{4N_\cmb + \alpha \sqrt{9 + 12/\alpha}}\right) 
\end{align}
in the T-model, and it is given by the solution of 
\begin{align}
	N_\cmb = \frac{3\alpha}{4} \biggl( e^{\sqrt{\frac{2}{3\alpha}}\frac{\phi_\cmb}{M_p}} + \ln\Bigl(1+\frac{2}{\sqrt{3\alpha}}\Bigr) - 1 \biggr) 
	- \frac{\sqrt{3\alpha}}{4} \biggl( 2 + \sqrt{2} \, \frac{\phi_\cmb}{M_p} \biggr) 
\end{align}
\end{subequations}
in the E-model.  
After accounting for the $\alpha$-dependence in $\phi_\cmb$, we see that $A_s$ is insensitive to $\alpha$ in the T-model.
In order to reproduce the value measured by \textit{Planck}, namely $\ln(10^{10} A_s) \simeq 3.044 \pm 0.014$~\cite{Akrami:2018odb}, we choose the scale parameter $\mu$ to satisfy \eref{eq:As} with $N_\cmb = 60$ for each \(\alpha\). 
After making this restriction, the T-model and E-model each have one free parameter, which is $\alpha$.  
For \(\alpha \ll N_\cmb\), the scalar spectral index, $n_s$, and the tensor-to-scalar ratio, $r$, are given by $1 - n_s \approx 2 / N_\cmb$ and $r \approx 12 \alpha / (N_\cmb (N_\cmb + 3 \alpha / 2))$~\cite{Kallosh:2015lwa}.  
A combo of \textit{Planck} and \textit{BICEP/Keck} infers an upper limit of $r < 0.064$ (95\% CL)~\cite{Akrami:2018odb}, which implies $\alpha \lesssim 38$ for \(N_\cmb = 60\).  

The energy scale of inflation is given by $H_\cmb \approx [V(\phi_\cmb) / 3 M_p^2]^{1/2}$ from \eref{eq:EOM}.  
Lowering $\alpha$ reduces the energy scale of inflation; for instance, the T-model has 
\begin{align}\label{eq:H_cmb}
	H_\cmb = \frac{4\pi \sqrt{6 \alpha A_s} M_p}{\sqrt{\bigl(4N_\cmb + \alpha \sqrt{9+12/\alpha}\bigr)^2 - 9\alpha^2}} 
	\com
\end{align}
which is approximately $H_\cmb \simeq (1.43 \times 10^{13} \GeV) (N_\cmb/60)^{-1} \sqrt{\alpha}$ for $\alpha \ll 2N_\cmb/3$. 

To close this section, let us comment on reheating.  
The inflaton condensate must eventually transfer its energy into a thermal bath of relativistic particles, thereby giving rise to the radiation-dominated universe, which connects to the standard Hot Big Bang cosmology~\cite{Kofman:1994rk}.  
Alpha attractor models admit several inflaton decay channels that can play a role in reheating~\cite{Ueno:2016dim,Eshaghi:2016kne,DiMarco:2017zek,Drewes:2017fmn,Krajewski:2018moi,Iarygina:2018kee,Iarygina:2020dwe}.  
We do not describe the physics of reheating in this work, but instead we assume that the universe remains matter-dominated during reheating, and that the radiation-dominated phase begins with a plasma temperature of $T_\RH$, which is taken to be a free parameter.  
However, we require that
\begin{align}\label{eq:low_TRH}
	T_\RH < \bigl( 8 \times 10^8 \GeV \bigr) \sqrt{ m_\chi / \GeV }
	\com
\end{align}
where $m_\chi$ is the mass of the scalar spectator field, corresponding to the gravitationally-produced dark matter.  
This condition ensures that $3H(t) < m_\chi$ before reheating is completed; \textit{e.g.}, see the ``late reheating'' regime of \rref{Kolb:2020fwh} (see also \rref{Ahmed:2020fhc}).  
In this regime, the comoving number density of gravitationally-produced particles is insensitive to the reheating temperature, and the late-time relic abundance only depends on $T_\RH$ through a simple scaling law \pref{eq:Omegah2}.  

\section{Gravitational particle production}\label{sec:GPP}

During inflation and reheating, fields that are coupled to gravity non-conformally, \textit{e.g.} through their mass, are put out of their vacuum state due to the nonadiabatic influence of the cosmological expansion.  
At late times, this excited state of the field corresponds to a nonzero density of particles that have been produced gravitationally.  
The subject of gravitational particle production has been studied extensively~\cite{Parker:1969au,Parker:1974qw,Fulling:1974zr,Ford:1986sy,Anderson:1987yt,Lyth:1996yj,Hashiba:2021npn}, and in the context of the inflaton field's quantum fluctuations, these are the seeds of structure that we observe on cosmological scales today.    
In the context of a spectator field, whose energy density is subdominant to the inflaton's during inflation, the phenomenon of gravitational particle production has important implications for the creation of superheavy particles, $m_\chi \sim H_\mathrm{inf}$, including a variety of dark matter candidates~\cite{Chung:1998zb,Kuzmin:1998uv,Giudice:1999yt,Kallosh:1999jj,Giudice:1999am,Dimopoulos:2006ms,Chung:2011ck,Graham:2015rva,Ema:2018ucl, Markkanen:2018gcw,Fairbairn:2018bsw,Hashiba:2018tbu,Guth:2018hsa,Ho:2019ayl,Tenkanen:2019aij,Herring:2019hbe,Ema:2019yrd,Hashiba:2019mzm,Herring:2020cah,Ahmed:2020fhc,Kolb:2020fwh}.  
In this section, we briefly review how one calculates the relic density of particles that results from gravitational particle production.  
For additional details on quantum fields in curved spacetime, we refer the reader to several excellent reviews and textbooks including Refs.~\cite{DeWitt:1975ys,BirrellDavies:1982,Parker:2009uva}.  

One can study the phenomenon of gravitational particle production during the inflationary era for various different theories, which are distinguished by the spectator field's representations under the Lorentz group, its mass, and its coupling to gravity.  
For concreteness, we focus here on the a real scalar spectator field, which corresponds to a spin-0 dark matter particle.  

Consider the real scalar field $\chi(x)$, whose properties and interactions are described by the action\footnote{If the inflaton has a large non-minimal coupling to gravity, then the spectator-field action may be divided by the conformal factor $\Omega = [1 + \xi \phi^2 / \Mpl^2]^{1/2}$, which can further enhance production of $\chi$ particles~\cite{Karam:2020rpa}.}  
\begin{align}\label{eq:00_action_1}
	S[\chi(x),\metric_{\mu \nu}(x)] = \int \! \ud^4 x \, \sqrt{-g} \, \biggl\{ 
	\frac{1}{2} \metric^{\mu \nu} \partial_{\mu} \chi \partial_{\nu} \chi 
	- \frac{1}{2} m_\chi^2 \chi^2 
	\biggr\} 
\end{align}
where $g = \mathrm{det}(g_{\mu\nu})$ is the metric determinant, and $\metric^{\mu\nu}$ is the inverse metric.
In the FRW spacetime \pref{eq:FRW}, the action becomes 
\begin{align}\label{eq:00_action_4}
	S[\chi(\ctime,\xvec)] = \int_{-\infty}^{\infty} \! \ud \ctime \int \! \ud^3 \xvec \, \biggl\{ 
	\frac{1}{2} \bigl[ \partial_\ctime (a\chi) \bigr]^2 
	- \frac{1}{2} \bigl[ {\bm \nabla} (a\chi) \bigr]^2 
	- \frac{1}{2} a^2 m_\eff^2 (a\chi)^2 
	- \frac{1}{2} \partial_\ctime \bigl[ aH (a\chi)^2 \bigr] 
	\biggr\} 
	\per
\end{align}
The last term is a total derivative, which doesn't contribute to the equations of motion, so we can neglect it.  
The field $a\chi$ has a canonically normalized kinetic term, and an effective squared mass parameter, $a^2 m_\eff^2$ with 
\begin{align}\label{eq:00_meff_def}
	m_\eff^2 \equiv 
	m_\chi^2 + \frac{1}{6} \, R 
	\com
\end{align}
which is time-dependent through both $a(\ctime)$ and $R(\ctime)$.  
For example, during inflation the spacetime metric is approximately the de Sitter one and we have $R \approx - 12 H^2$, which gives $m_\eff^2 \approx m_\chi^2 - 2H^2$.  
For a sufficiently light scalar spectator (small $m_\chi$) the field may be tachyonic ($m_\eff^2 < 0$).  

Varying the action \pref{eq:00_action_4} with respect to the field yields the field equation, 
\begin{align}\label{eq:00_field_eqn_3}
	\bigl[ \partial_\ctime^2 - {\bm \nabla}^2 + a^2 m_\eff^2 \bigr] (a \chi) = 0 
	\per
\end{align}
When expressed in terms of conformal time, $\ctime$, and the canonically-normalized field, $a\chi$, this field equation is simply the usual Klein-Gordon one with the replacement $m_\chi^2 \to a^2 m_\eff^2$.  
Here we can anticipate how the cosmological expansion must lead to particle production.  
If there were no cosmological expansion ($a=1$ and $R = 0$) then \eref{eq:00_field_eqn_3} would reduce to the familiar Klein-Gordon equation, and its solutions would be the usual plane waves.  
In this regime, no particle production occurs, since the vacuum mode functions have a static amplitude.  
However, if the universe is expanding ($\partial_\ctime a \neq 0$), then the vacuum mode functions develop time-dependent amplitudes, as the field ``responds'' to its varying effective mass, and there is a particle production associated with the nonadiabaticity of this mass evolution.  

Since the inflationary spacetime is nearly homogeneous, and described by the FRW metric, it is convenient to perform a Fourier decomposition of the field.  
We introduce a set of complex-valued mode functions $\chi_\kvec(\ctime)$ labeled by a 3-vector $\kvec$ called the comoving wavevector.  
The Fourier decomposition is written as 
\begin{align}
	\chi(\ctime,\xvec) & = a(\ctime)^{-1} \int \! \! \frac{\ud^3 \kvec}{(2\pi)^3} \, \chi_\kvec(\ctime) \, e^{i \kvec \cdot \xvec} 
	\com
\end{align}
where the mode functions obey $\chi_\kvec(\ctime) = \chi_{-\kvec}(\ctime)^\ast$, ensuring that $\chi(\ctime,\xvec) = \chi(\ctime,\xvec)^\ast$ is real.  
The field equation \pref{eq:00_field_eqn_3} yields a set of equations of motion for the various mode functions 
\begin{align}\label{eq:00_field_eqn_3b}
	\bigl( \partial_\ctime^2 + \omega_k^2 \bigr) \chi_\kvec = 0 
\end{align}
where we have defined the comoving squared angular frequency, 
\begin{align}\label{eq:00_omega}
	\omega_k^2 = k^2 + a^2 m_\chi^2 + \frac{1}{6} a^2 R 
	\com
\end{align}
which only depends on $\kvec$ via the comoving wavenumber $k = |\kvec|$.  

The mode equations \pref{eq:00_field_eqn_3b} must be solved along with a set of initial conditions.  
To study gravitational particle production in the inflationary era, we recognize that the early-time limit $\ctime \to -\infty$ corresponds to $a \to 0$, $a^2 R \to 0$, and $\omega_k \to k$.  
In other words, as we send $\ctime \to -\infty$ the modes with $|\kvec| = k$ are deep inside the horizon, meaning that $aH \ll k$, and they are relativistic, meaning that $am_\chi \ll k$.  
Consequently, their mode equation is approximately the same as the one in Minkowski space with $\omega_k = k$.  
This observation motivates the so-called Bunch-Davies initial condition, 
\begin{align}\label{eq:00_Bunch_Davies}
	\lim_{\ctime \to - \infty} \chi_\kvec(\ctime) = \frac{1}{\sqrt{2k}} \, e^{-i k \ctime} 
	\com
\end{align}
where the factor of $1/\sqrt{2k}$ ensures that the canonical commutation relations are properly normalized.  

For a given spacetime background, encoded in $a(\ctime)$, and for a given model, parametrized by $m_\chi^2$, the mode equations \pref{eq:00_field_eqn_3b} are solved along with the corresponding Bunch-Davies initial condition \pref{eq:00_Bunch_Davies} to obtain the mode functions $\chi_\kvec(\ctime)$.  
The spectrum of gravitationally produced particles is derived from the Bogoliubov coefficients, $\alpha_\kvec$ and $\beta_\kvec$, that relate an observer at asymptotically early time to an observer at asymptotically late time.\footnote{For a pedagogical discussion of Bogoliugov coefficients in an inflationary context, we refer the reader to Refs.~\cite{BirrellDavies:1982,Mukhanov:2005,Baumann:2009ds,Parker:2009uva}.}
The relevant combinations of Bogoliubov coefficients are extracted using 
\begin{subequations}\label{eq:identities}
\begin{align}
	\label{eq:beta_k_sq}
	|\beta_\kvec|^2 & = \lim_{\ctime \to \infty} \biggl[ \frac{\omega_k}{2} \, |\chi_\kvec|^2 + \frac{1}{2 \omega_k} \, |\partial_\ctime \chi_\kvec|^2 + \frac{i}{2} \bigl( \chi_\kvec \, \partial_\ctime \chi_\kvec^\ast - \chi_\kvec^\ast \partial_\ctime \chi_\kvec \bigr) \biggr] \\
	\mathrm{Re}\bigl[ \alpha_\kvec \beta_\kvec^\ast e^{-2i \int^\ctime \! \omega_k} \bigr] & = \lim_{\ctime \to \infty} \biggl[ \frac{\omega_k}{2} \, |\chi_\kvec|^2 - \frac{1}{2 \omega_k} \, |\partial_\ctime \chi_\kvec|^2 \biggr] \\ 
	\mathrm{Im}\bigl[ \alpha_\kvec \beta_\kvec^\ast e^{-2i \int^\ctime \! \omega_k} \bigr] & = \lim_{\ctime \to \infty} \biggl[ \frac{1}{2} \bigl( \chi_\kvec \, \partial_\ctime \chi_\kvec^\ast + \chi_\kvec^\ast \partial_\ctime \chi_\kvec \bigr) \biggr] 
\end{align}
\end{subequations}
where $\int^\ctime \! \omega_k \equiv \int_{-\infty}^\ctime \! \ud \ctime^\prime \, \omega_k(\ctime^\prime)$.  
Thanks to the spatial isotropy, $|\beta_\kvec|^2 = |\beta_k|^2$ only depends on $k = |\kvec|$.  

The spectrum of gravitationally-produced particles measured by the late-time observer is calculated from $|\beta_k|^2$.  
The comoving number density of $\chi$ particles is\footnote{See, for example, Secs.~2.8 and 2.9 of \rref{Parker:2009uva}.}
\begin{align}\label{eq:a3n}
	a^3 n = \int_{k_0}^\infty \frac{\ud k}{k} \ \Ncal_k
\end{align}
where the comoving number density spectrum is
\begin{align}\label{eq:Nk_from_betaksq}
	\Ncal_k 
	= a^3 \, k \frac{dn}{dk}
	= \frac{k^3}{2\pi^2} \, |\beta_k|^2 
	\per
\end{align}
Note that $\Ncal_k$ is static (independent of $\ctime$) and the comoving number density is conserved, i.e. $\partial_\ctime(a^3n) = 0$.  
Similarly, the comoving energy density is\footnote{In the derivation of these expressions for $a^3 n$ and $a^3 \rho$, one encounter ultraviolet (UV) divergences that must be regulated and renormalized.  We have used the standard adiabatic regularization scheme~\cite{Parker:1974qw,Ford:1986sy,Fulling:1974zr,Anderson:1987yt}.  So $n$ and $\rho$ correspond to the renormalized quantities.}  
\begin{align}
	a^3 \rho = \int_{k_0}^\infty \frac{\ud k}{k} \ \Ecal_k(\ctime) 
	\qquad \text{with} \qquad 
	\Ecal_k(\ctime) = \frac{k^3}{2\pi^2} \, \frac{\omega_k(\ctime)}{a(\ctime)} \, |\beta_k|^2 
\end{align}
where $\omega_k/a \approx [(k/a)^2 + m_\chi^2]^{1/2}$ up to $O(R, H^2)$ terms.  
For non-relativistic modes, $\Ecal_k = m_\chi \, \Ncal_k$.  
We introduce an IR cutoff, $k_0 = a_0 H_0$, corresponding to the comoving wavenumber of modes entering the horizon today, $a_0 \equiv a(\ctime_0)$ and $H_0 \equiv H(\ctime_0)$.  
For $m_\chi \lesssim H_e$ the spectra are slightly red-tilted and $k_0$ regulates the integrals, but for $m_\chi \gtrsim H_e$, which is the regime of interest, the low-$k$ spectra are blue-tilted and the integrals are insensitive to $k_0$.  
Modes with $k < k_0$ are effectively homogeneous on the scale of our Hubble patch and their incoherent superposition renormalizes the zero mode $k = 0$; see also Refs.~\cite{Bunch:1978yq,Linde:1982uu,Starobinsky:1994bd}.  

If the gravitationally-produced particles are sufficiently long lived, they can survive in the universe today, at conformal time $\ctime = \ctime_0$, thereby providing a candidate for the dark matter.  
We quantify the abundance of these particles with the comoving number density, $a^3 n$ from \eref{eq:a3n}, since this quantity is conserved in the absence of particle-number-changing interactions, and we have $a^3 n = a_0^3 n_0$ where $a_0 \equiv a(\ctime_0)$ and $n_0 \equiv n(\ctime_0)$.  
Alternatively we can express their density today as a relic abundance, $\Omega h^2 = \rho_0 / (3 M_p^2 H_{100}^2)$ where $\rho_0 \equiv \rho(\ctime_0) \approx m_\chi n_0$ is their energy density today, $M_p \simeq 2.435 \times 10^{18} \GeV$ is the reduced Planck mass, and $H_{100} \equiv 100 \km / \mathrm{sec} / \mathrm{Mpc}$.  
Assuming that the comoving entropy density of the primordial plasma is conserved, we can express the relic abundance as~\cite{Kolb:2020fwh} 
\begin{equation}\label{eq:Omegah2}
\begin{split}
	\Omega h^2 
	& = \biggl( \frac{\pi^2 g_{\ast S,0} T_0^3}{270 M_p H_{100}^2} \biggr) 
	\biggl( \frac{m_\chi H_e T_\RH}{M_p^3} \biggr)
	\biggl( \frac{a^3 n}{a_e^3 H_e^3} \biggr) \\ 
	& \simeq 
	\bigl( 0.114 \bigr)
	\biggl( \frac{m_\chi}{10^{10} \GeV} \biggr)
	\biggl( \frac{H_e}{10^{10} \GeV} \biggr)
	\biggl( \frac{T_\RH}{10^{8} \GeV} \biggr)
	\biggl( \frac{a^3 n}{a_e^3 H_e^3} \biggr) 
\end{split}
\end{equation}
where $g_{\ast S,0} \simeq 3.91$ and $T_0 \simeq 0.234 \meV$.  
The measured relic abundance of dark matter is $\Omega_\mathrm{dm} h^2 \simeq 0.12$~\cite{Aghanim:2018eyx}.  

The isocurvature power spectrum in the comoving gauge is 
\begin{align}\label{eq:isocurvature_def}
	\Delta_\Scal^2(\ctime,k) & = \frac{1}{\rho(\ctime)^2} \, \frac{k^3}{2\pi^2} \int \! \ud^3 {\bm r} \, \langle \drho(\ctime,\xvec) \, \drho(\ctime,\xvec+{\bm r}) \rangle \, e^{-i \kvec \cdot {\bm r}} 
\end{align}
where the energy density in the $\chi$ field at time $\ctime$ has been written as $\rho(\ctime) + \drho(\ctime,\xvec)$.  
By expressing $\drho(\ctime,\xvec)$ in terms of $\chi_\kvec(\ctime)$, we can write~\cite{Chung:2004nh,Chung:2011xd}  
\begin{equation}
\begin{split}
	\Delta_\Scal^2(k) 
	& = \frac{a^{-8}}{2\rho^2} \frac{k^2}{2\pi^2} \int \! \! \frac{\ud^3 \kvec^\prime}{(2\pi)^3} \biggl\{ 
	|\partial_\ctime \chi_{\kvec^\prime}|^2 \, |\partial_\ctime \chi_{\kvec-\kvec^\prime}|^2 
	+ a^4 m^4 \, |\chi_{\kvec^\prime}|^2 \, |\chi_{\kvec-\kvec^\prime}|^2 
	\\ & \qquad 
	+ a^2 m^2 \Bigl[ (\chi_{\kvec^\prime} \partial_\ctime \chi_{\kvec^\prime}^\ast) (\chi_{\kvec-\kvec^\prime} \partial_\ctime \chi_{\kvec-\kvec^\prime}^\ast) + (\chi_{\kvec^\prime}^\ast \partial_\ctime \chi_{\kvec^\prime}) (\chi_{\kvec-\kvec^\prime}^\ast \partial_\ctime \chi_{\kvec-\kvec^\prime}) \Bigr] 
	\biggr\} 
	\com
\end{split}
\end{equation}
and by using the identities in \eref{eq:identities}, we can also write this as 
\begin{align}\label{eq:isocurvature}
	\Delta_\Scal^2(k) 
	= \frac{a^{-3}}{n(\ctime)} \frac{k^3}{2\pi^2} 
	+ \frac{a^{-6}}{n(\ctime)^2} \frac{k^3}{2\pi^2}
	& 
	\int \! \! \frac{\ud^3 \kvec^\prime}{(2\pi)^3} \biggl\{ 
	|\beta_{\kvec^\prime}|^2 \, |\beta_{\kvec-\kvec^\prime}|^2 
	\\ & 
	+ \mathrm{Re}\Bigl[ \alpha_{\kvec^\prime} \beta_{\kvec^\prime}^\ast e^{-2i\int^\ctime \omega_{\kvec^\prime}} \Bigr] \, \mathrm{Re}\Bigl[ \alpha_{\kvec-\kvec^\prime} \beta_{\kvec-\kvec^\prime}^\ast e^{-2i\int^\ctime \omega_{\kvec-\kvec^\prime}} \Bigr] 
	\nn & 
	+ \mathrm{Im}\Bigl[ \alpha_{\kvec^\prime} \beta_{\kvec^\prime}^\ast e^{-2i\int^\ctime \omega_{\kvec^\prime}} \Bigr] \, \mathrm{Im}\Bigl[ \alpha_{\kvec-\kvec^\prime} \beta_{\kvec-\kvec^\prime}^\ast e^{-2i\int^\ctime \omega_{\kvec-\kvec^\prime}} \Bigr] 
	\biggr\} \nonumber 
	\per
\end{align}
The first term $\propto k^3$ is entirely negligible for the CMB-scale modes, and generally only the second term will be relevant.   

\section{Numerical results}\label{sec:DM}

In this section we present the main results of our study, including the evolution of the scalar spectator field's mode functions in an $\alpha$-attractor model of inflation, the spectrum of gravitationally-produced particles, the isocurvature power spectrum, and the constraints imposed by \textit{Planck} data.  

\subsection{Evolution}\label{sub:evolution}

\begin{figure}[t!]
\begin{center}
\includegraphics[width=0.49\textwidth]{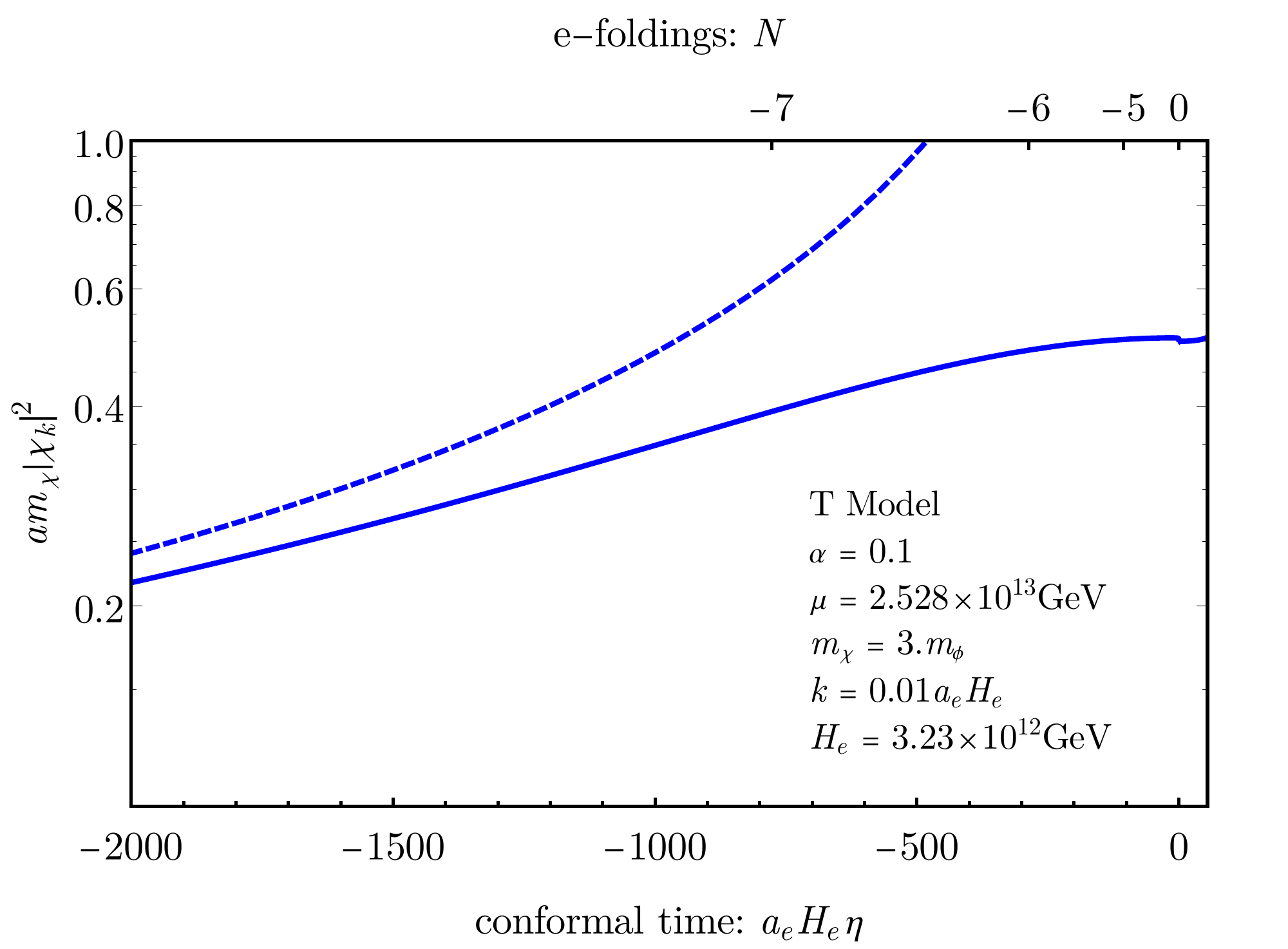} 
\includegraphics[width=0.49\textwidth]{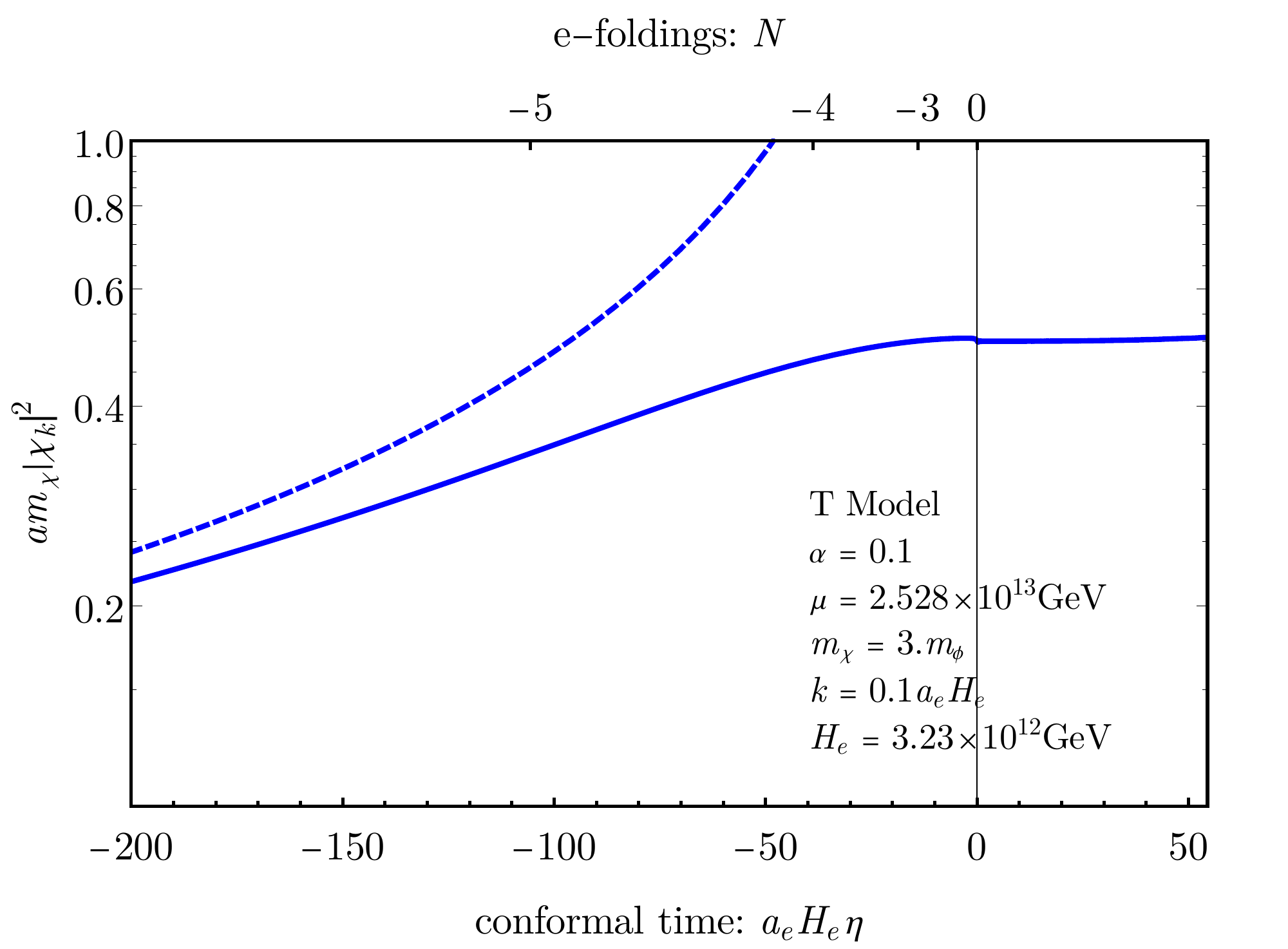} \\
\includegraphics[width=0.49\textwidth]{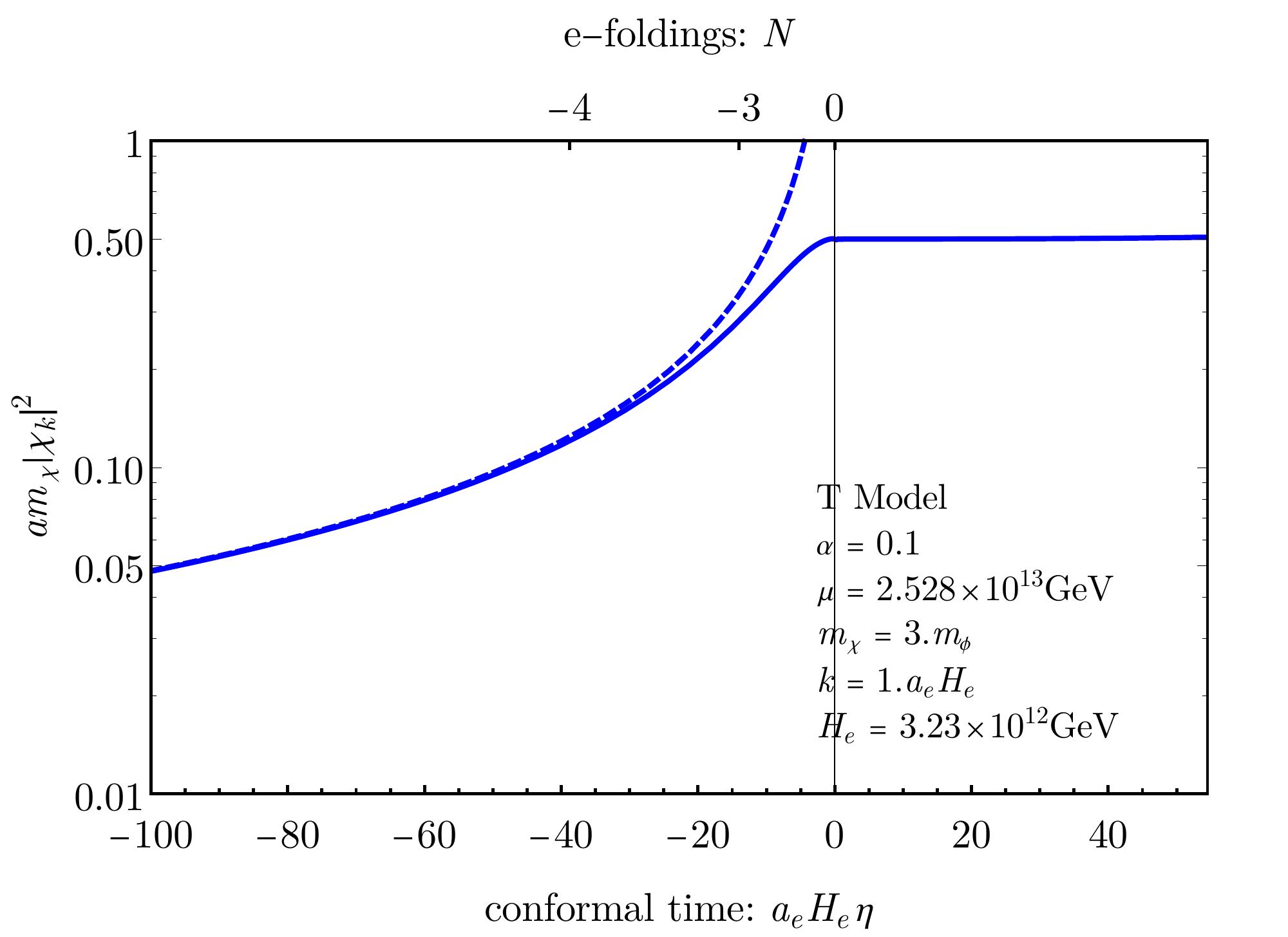} 
\includegraphics[width=0.49\textwidth]{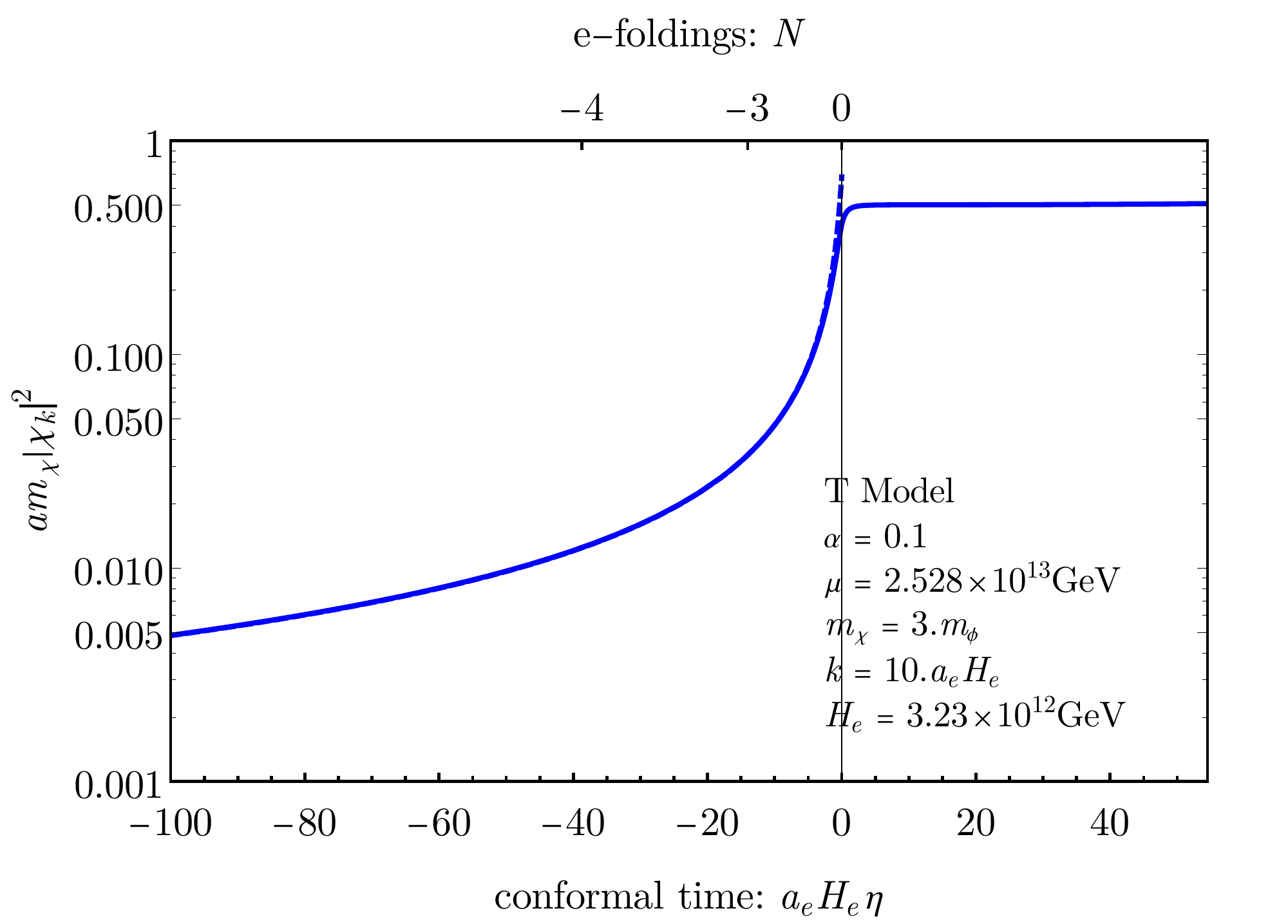} 
\caption{\label{fig:evolution}\small
Evolution of the scalar spectator's mode functions near the end of inflation.  All panels correspond to the T-model $\alpha$-attractor with $\alpha = 0.1$ and $\mu = 2.528 \times 10^{13} \GeV$ and a minimally-coupled scalar spectator with mass $m_\chi = 3 m_\phi$.  From top to bottom the rows correspond to different modes with wavenumbers $k / a_e H_e = 0.01$, $0.1$, $1$, and $10$.  For each mode we show the scaled amplitude of the mode function, $a(\ctime) m_\chi |\chi_\kvec(\ctime)|^2$, as a function of scaled conformal time $a_e H_e \ctime$.  Inflation ends at $\ctime = \ctime_e = 0$.  The solid-blue curve shows the numerical solution, and the dashed-blue curve shows the Bunch-Davies mode function \pref{eq:00_Bunch_Davies}, which sets the initial condition at early time.  
}
\end{center}
\end{figure}

The mode equations \pref{eq:00_field_eqn_3b} are solved numerically for a few choices of $k = |\kvec|$ and model parameters, and we present the resulting mode functions in \fref{fig:evolution}.  
Initially $\chi_\kvec(\ctime)$ matches the Bunch-Davies vacuum mode function \pref{eq:00_Bunch_Davies}, indicated by the dashed curve.  
As inflation continues, $\chi_\kvec(\ctime)$ starts to deviate from the Bunch-Davies vacuum; this effect is more pronounced for smaller $k/a_e H_e$, corresponding to the first two panels.  
After inflaton ends, the mode functions begin to oscillate in such a way that $a |\chi_\kvec|^2$ has a fixed amplitude.  
At late times when the mode of interest is inside the horizon ($aH < k$) and nonrelativistic ($am_\chi < k$), the dispersion relation is approximately $\omega_k \approx am_\chi$, and the quantity plotted here, $a m_\chi |\chi_\kvec|^2$ is just the first term in the Bogoliubov coefficient, $|\beta_\kvec|^2$ from \eref{eq:beta_k_sq}.  

We can understand the evolution of these mode functions by solving the mode equations analytically in different regimes of interest.  
See Refs.~\cite{Ahmed:2020fhc,Kolb:2020fwh} for a similar analysis in the case of a vector spectator field.  
Combining \erefs{eq:00_field_eqn_3b}{eq:00_omega} lets us write the mode equations as 
\begin{align}
	\Bigl( \partial_\ctime^2 + k^2 + a^2 m_\chi^2 + \frac{1}{6} a^2 R \Bigr) \chi_\kvec = 0 
	\per
\end{align}
During inflation the Ricci scalar is $R \approx -12 H^2$ and the Hubble parameter is approximately constant $H \approx H_e$, implying $a(\ctime) \approx a_e / [ 1 - a_e H_e (\ctime - \ctime_e)]$.  
Long before the end of inflation, $a \approx - 1 / (H_e \ctime)$.  
In \fref{fig:evolution} we have shown only a model with $m_\chi = 3 m_\phi$.  
For such a model, the mode equation is approximately 
\begin{align}
	\Bigl[ \partial_\ctime^2 + k^2 + \frac{1}{\ctime^2} \Bigl( \frac{m_\chi^2}{H_e^2} - 2 \Bigr) \Bigr] \chi_\kvec = 0 
	\qquad \text{(during inflation, $\ctime < \ctime_e = 0$)}
	\per
\end{align}
Here, \(m_\chi^2 / H_e^2 - 2\) is a positive number of order unity.
Initially the mode in question is deep inside the horizon, $k^2 \gg 1/\ctime^2$, and the solution is $\chi_\kvec \sim e^{-i k \ctime}$ giving $a m_\chi |\chi_\kvec|^2 \sim a$.  
For $k < a_e H_e$ the mode in question will leave the horizon before the end of inflation.  
Once the mode is outside the horizon, $k^2 \ll 2/\ctime^2$, the solution becomes $\chi_\kvec \sim \ctime^{1/2} \sim a^{-1/2}$ and $am_\chi |\chi_\kvec|^2 \sim a^0$.  
After inflation is ended, the non-adiabaticity drops far below unity, the solution becomes \(\chi_\kvec \sim a^{-1/2}\) and \(a m_\chi |\chi_\kvec|^2\) converges to a constant.
We can see from \fref{fig:evolution} that the numerical solution \(a m_\chi |\chi_\kvec|^2\) tracks the Bunch-Davies initial condition at early times and converges to a constant after inflation ends.
Note that for modes with higher \(k\), the numerical solution tracks the Bunch-Davies initial condition to until later times in the inflation; this highlights the fact that lower-\(k\) modes leave the horizon earlier, as they should due to the horizon-crossing condition \(k = a H\).

\subsection{Spectrum}\label{sub:spectrum}

\begin{figure}[t!]
\begin{center}
\includegraphics[width=0.49\textwidth]{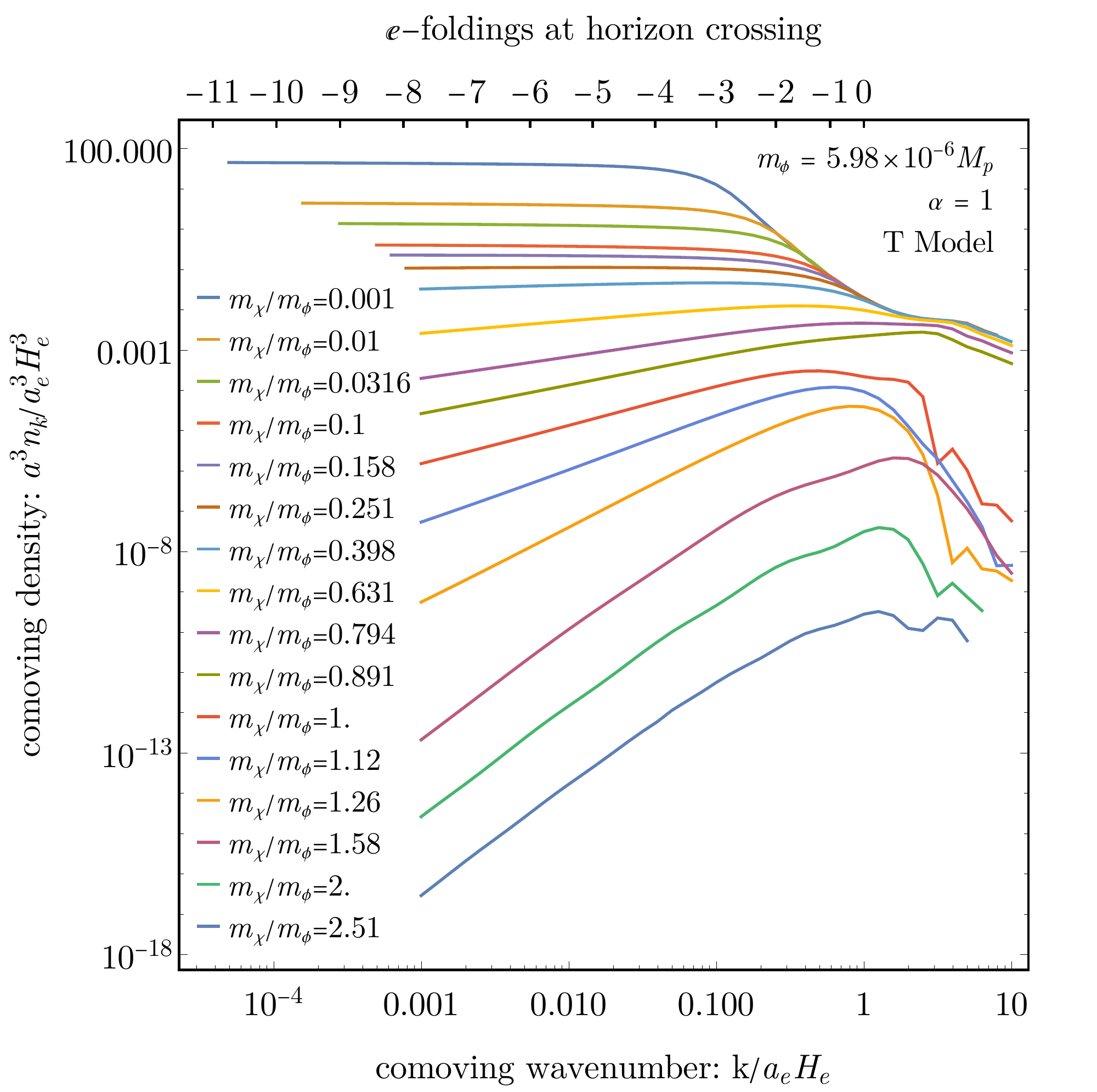} 
\includegraphics[width=0.49\textwidth]{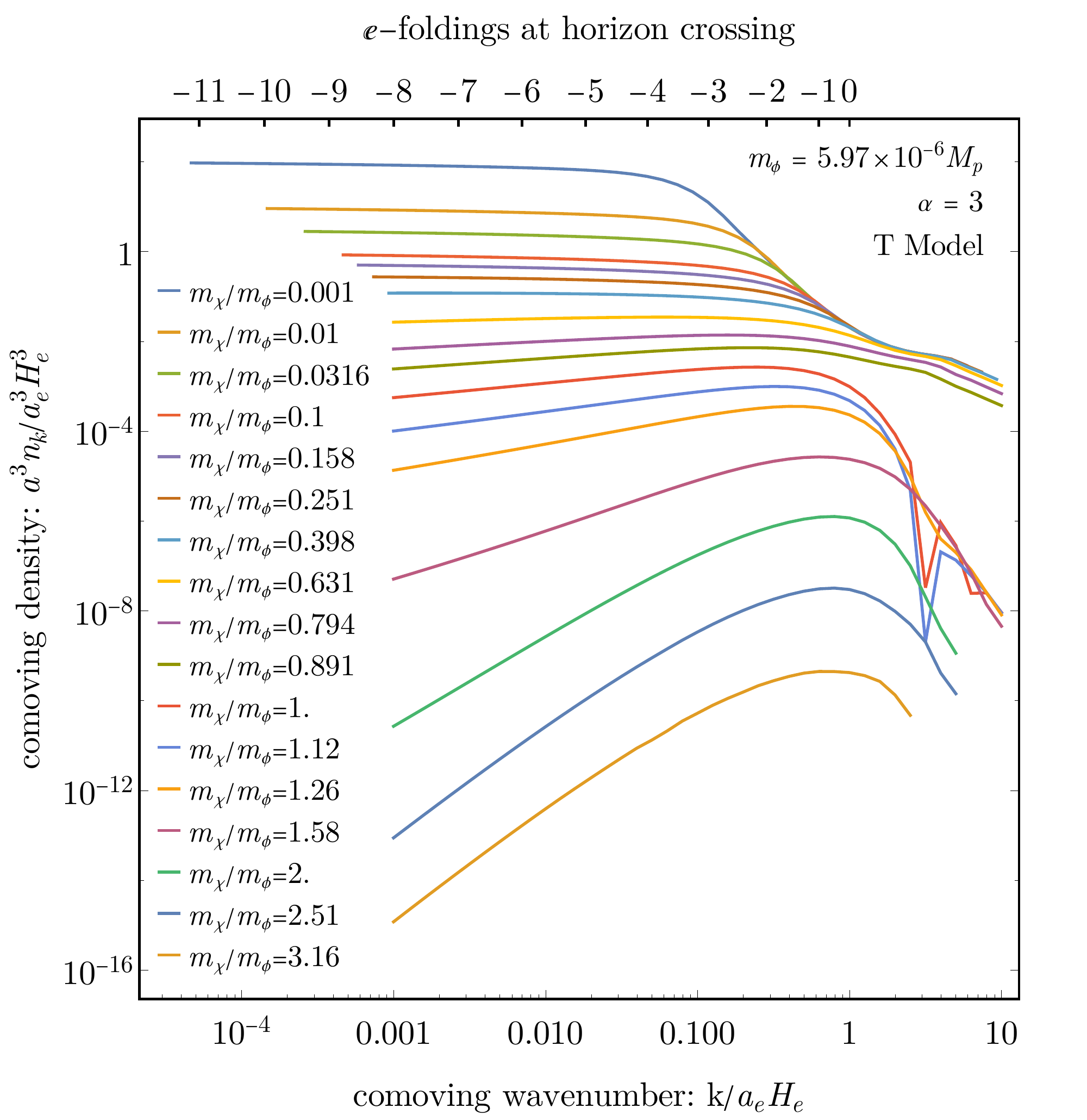} \\ 
\includegraphics[width=0.49\textwidth]{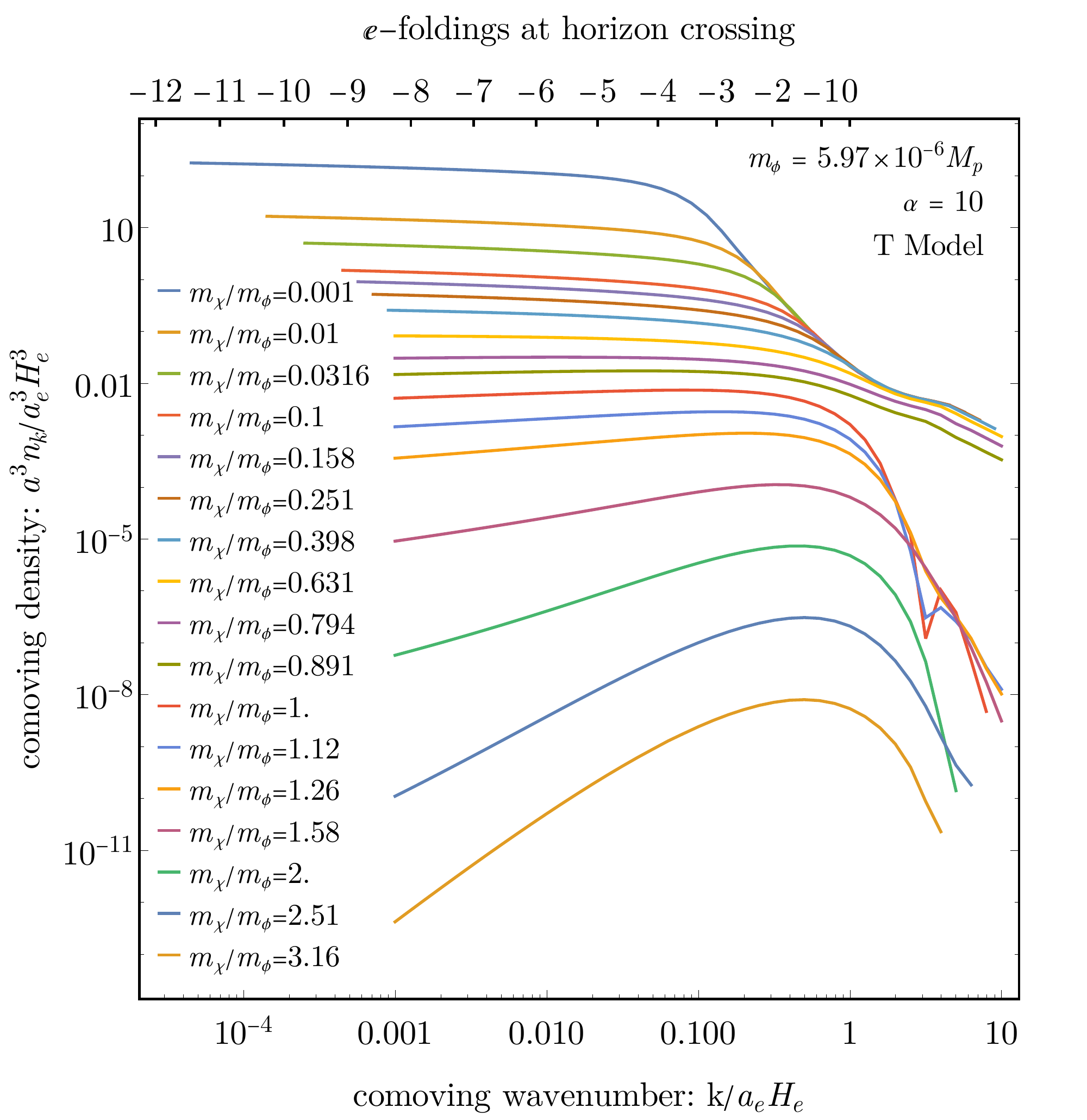} 
\includegraphics[width=0.49\textwidth]{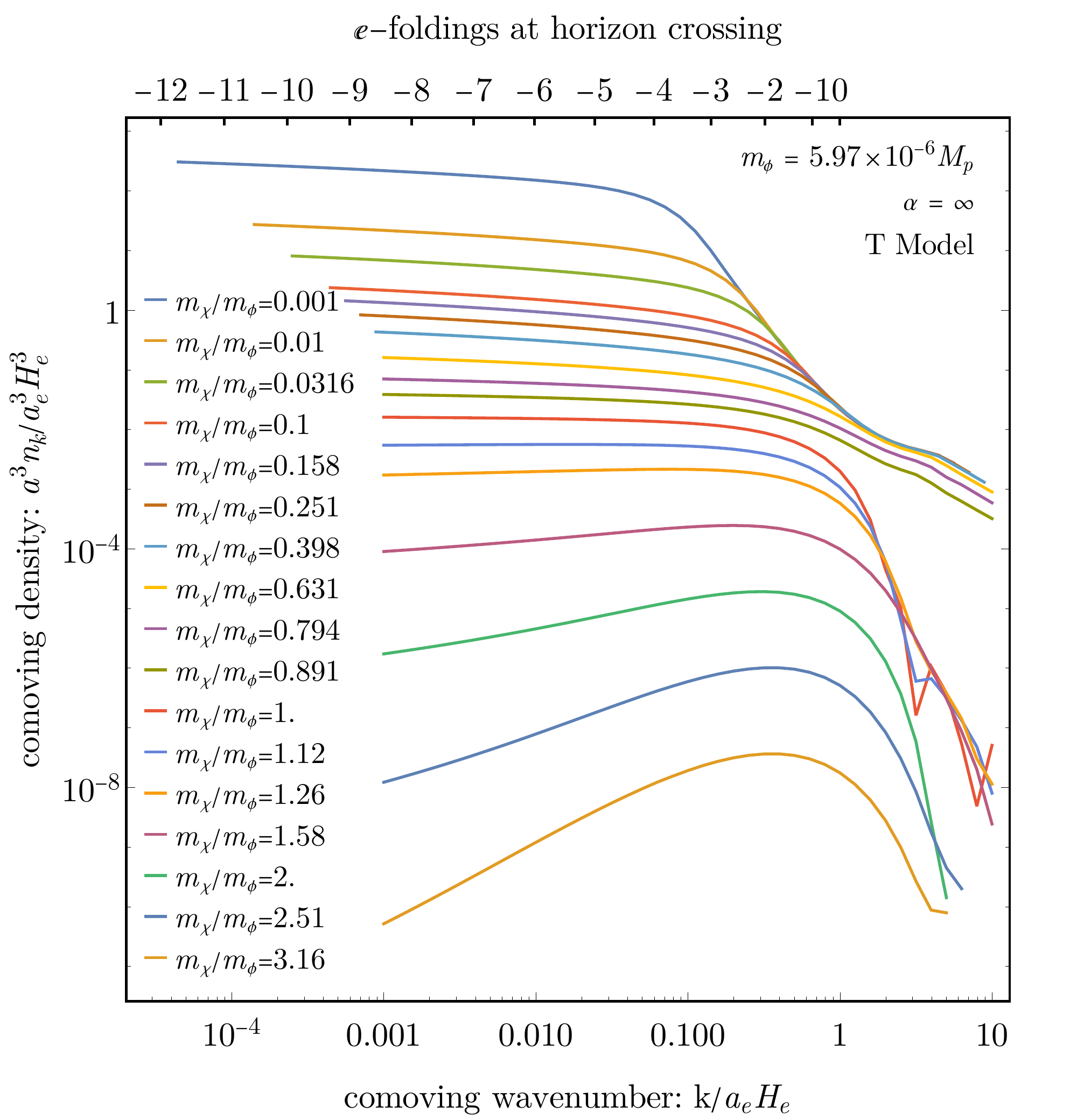} 
\caption{\label{fig:spectrum}\small
The spectrum of spin-0 particles arising from gravitational particle production in an $\alpha$-attractor model of inflation.  We show the comoving number spectrum $a^3 n_k$ where $n_k \equiv k \, dn/dk$ in units of $a_e^3 H_e^3$.  All four panels correspond to the T-model $\alpha$-attractor with inflaton mass $m_\phi \approx 6 \times 10^{-6} M_p$, and the value of $\alpha$ is varied across the panels from $\alpha = 1$, $3$, $10$, to $\infty$.  In each panel, the various curves show different values of the scalar spectator's mass from $m_\chi = 0.001 \, m_\phi$ to $3.16 \, m_\phi$.  
}
\end{center}
\end{figure}

The spectrum of gravitationally-produced particles is calculated by scanning over a range of $k = |\kvec|$, and the results are shown in \fref{fig:spectrum}.  
Recall from \eref{eq:Nk_from_betaksq} that $\Ncal_k = a^3 \, k \, dn/dk = k^3 |\beta_k|^2/2\pi^2$ is the comoving number density per logarithmic wavenumber interval.  
In particular, we are interested in how the spectrum changes as $\alpha$ is varied.  

For $\alpha \gg 1$ we regain the well-known results for chaotic inflation with $V = m_\phi^2\phi^2/2$~\cite{Kuzmin:1998kk}.  
In particular, for a light spectator ($m_\chi \ll H_e$) the spectrum is nearly scale invariant for modes that left the horizon during inflation ($k \ll a_e H_e$), with the scale invariance-violation induced by the gradual decrease of the Hubble parameter ($\partial_\ctime H < 0$).  
For modes that remained inside the horizon throughout inflation ($k > a_e H_e$) the spectrum is highly suppressed~\cite{Chung:1998bt,Chung:1998zb,Chung:1998ua,Kuzmin:1998kk,Chung:2001cb}.  
Raising the mass of the scalar spectator tilts the slightly-red spectrum into a blue spectrum, meaning that $\Ncal_k \to 0$ as $k \to 0$, which is a familiar result from the study of light spectators in de Sitter spacetime~\cite{Mukhanov:2005,Baumann:2009ds}.  

For smaller values of $\alpha$, the energy scale of inflation decreases and the plateau in the $\alpha$-attractor inflaton potential becomes more relevant; see \fref{fig:potential}.  
By lowering the energy scale of inflation with smaller $\alpha$, the abundance of gravitationally-produced particles is reduced, and this is reflected in smaller values of the spectra $a^3 \, k \, dn/dk$.
More specifically, numerical results suggest that \(a^3 \, k \, dn/dk \sim H_e^3\) holds at low-\(\alpha\) and low-\(k\),
and \(H_e \sim \alpha^{1/2}\) holds at low-\(\alpha\).
The scaling of \(H_e\) can be understood in the following way:
\(H_e\) can be estimated via \(H_e \approx [V(\phi_f) / 3 M_p^2]^{1/2}\),
where \(\phi_f\) is set by the end of inflation condition in terms of the potential \(M_p^2 (V'(\phi_f) / V(\phi_f))^2 / 2 = 1\).
For the T-model, this gives us:
\begin{align}
	H_e \approx \sqrt{\frac{\alpha}{3}} \, \mu \tanh \biggl( \frac{1}{2} \sinh^{-1}\Bigl( \frac{2}{\sqrt{3\alpha}} \Bigr) \biggr)
	\per
\end{align}
It is clear from this estimate that \(H_e \sim \alpha^{1/2}\) at low-\(\alpha\).
Numerical results also confirm that this estimate is accurate up to \(O(0.1)\).
Note that as we lower \(\alpha\), the ratio \(H_\cmb / H_e\) converges to \(1\).
This can be seen as an effect due to the flattening of the potential \(V\):
as \(V\) becomes flatter, the hubble scale experience less change during inflation;
see \fref{fig:potential}.

In the high-momentum tail of the spectrum ($k \gtrsim a_e H_e$) a threshold effect can be seen.  
If the mass of the scalar spectator is smaller than the mass of the inflaton, $m_\chi \leq m_\phi$, then the high-$k$ spectrum drops off like a power law from $k/a_e H_e = 1$ to $10$.  
However, if the scalar's mass is larger, $m_\chi > m_\phi$, then the spectrum drops more steeply, as an exponential.  
This abrupt change in behavior at $m_\chi = m_\phi$ can be understood as a threshold effect~\cite{Ema:2018ucl,Chung:2018ayg}, as we now explain.  
When \(m_\chi \leq m_\phi\), the high-\(k\) particles are primarily produced by oscillations in the inflaton field after the end of inflation.
It is helpful to understand this kind of particle production as a \(\phi\phi\to\chi\chi\) scattering process, mediated by an off-shell graviton~\cite{Ema:2018ucl}.  
By conservation of energy, we know that such a process cannot happen at \(m_\chi > m_\phi\).
Moreover, in the case that \(m_\chi \leq m_\phi\), 
it is possible to derive a formula for the high-\(k\) tail of the spectrum, and show that the spectral index of this tail is \(-3/2\).

\subsection{Relic abundance}\label{sub:abundance}

We calculate the comoving number density of gravitationally-produced particles, $a^3 n$, by integrating the spectrum, $\Ncal_k$, as in \eref{eq:a3n}.  
Then the relic abundance, $\Omega h^2$, is given by \eref{eq:Omegah2}.  
Recall that there are four free model parameters: $\{ \text{T- or E-model}, \alpha, T_\RH, m_\chi \}$ where the reheating temperature $T_\RH$ only enters $\Omega h^2$ as a simple scaling law.  
Rather than performing an exhaustive parameter-space scan, we show several representative ``slices'' of this parameter space to illustrate the parametric dependence.  

\begin{figure}[t!]
\begin{center}
\includegraphics[width=0.49\textwidth]{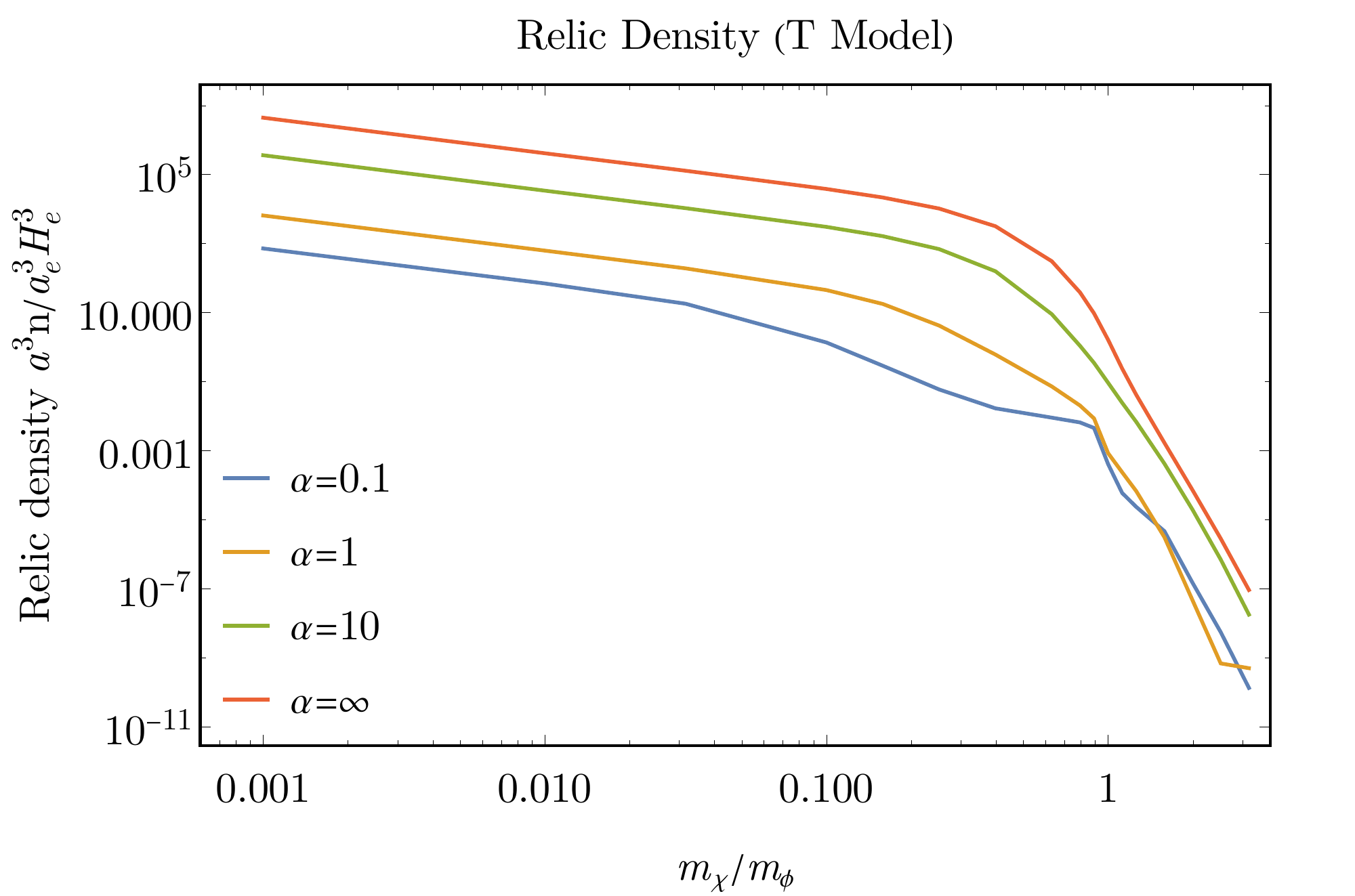} \hfill
\includegraphics[width=0.49\textwidth]{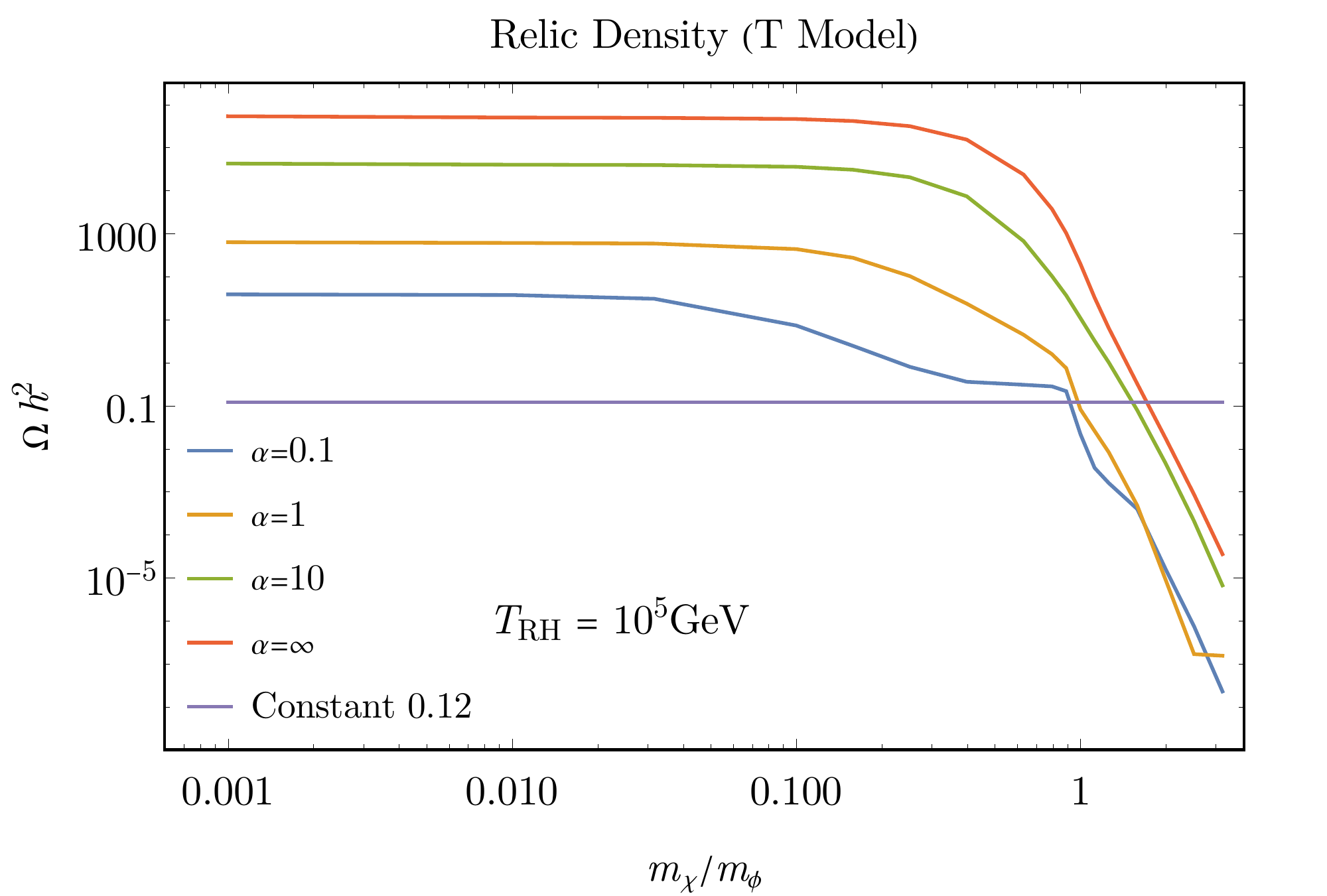} 
\caption{\label{fig:abundance_1}\small
The abundance of spin-0 particles arising from gravitational particle production in $\alpha$-attractor models of inflation.  The left panel shows the comoving number density $a^3 n$ in units of $a_e^3 H_e^3$, and the right panel shows the dimensionless relic abundance $\Omega h^2$.  This calculation is performed for the T-model $\alpha$-attractor with $\mu = 2.5 \times 10^{13} \GeV$ and $m_\phi = 1.5 \times 10^{13} \GeV$.  We vary $\alpha$ and the spectator's mass, $m_\chi$.  The right panel also takes $T_\RH = 10^5 \GeV$, and for models with a different reheating temperature, the relic abundance can be inferred from \eref{eq:Omegah2} and using the left panel.  
}
\end{center}
\end{figure}

First we consider the T-model $\alpha$-attractor.  
The comoving number density and relic abundance are shown in \fref{fig:abundance_1} as a function of the parameter $\alpha$ and the scalar spectator's mass $m_\chi$.  
For a given value of $\alpha$ the mass dependence follows the expected behavior.  
For a light spectator ($m_\chi \ll H_e$), the comoving density goes as $a^3 n \propto m_\chi^{-1}$ and the relic abundance is insensitive to the mass, going as $\Omega h^2 \propto m_\chi^0$.  
For a heavy spectator ($H_e \lesssim m_\chi$), we see that gravitational production is suppressed~\cite{Chung:1998bt}.  

It is interesting to see how the amount of gravitational particle production changes as $\alpha$ is varied.
\eref{eq:Omegah2} tells us that the relic abundance \(\Omega h^2\) is proportional to the product of three factors: \(a^3 n / a_e^3 H_e^3\), \(m_\chi\) and \(H_e\).
By fixing a ratio \(m_\chi / H_e\), we can eliminate the free parameter \(m_\chi\)
and conclude that \(\Omega h^2\) is proportional to \((a^3 n / a_e^3 H_e^3) H_e^2\),
leaving only dependence on a single parameter \(\alpha\).
For \(\alpha \gg 1\) we regain the familiar results for \(a^3 n / a_e^3 H_e^3\) and \(H_e\)
in the case of chaotic inflation with \(V = m_\phi^2 \phi^2\)~\cite{Kuzmin:1998kk}.
Since both \(a^3 n / a_e^3 H_e^3\) and \(H_e\) converge as \(\alpha \to \infty\),
we have \(\Omega h^2 \sim \alpha^0\) in that limit.
On the other hand, for low \(\alpha\) we know from our discussion in \sref{sub:spectrum} that
\(a^3 n / a_e^3 H_e^3 \sim \alpha^0\) and \(H_e \sim \alpha^{1/2}\),
so we have \(\Omega h^2 \sim \alpha^1\).

\begin{figure}[t]
\begin{center}
\includegraphics[width=0.55\textwidth]{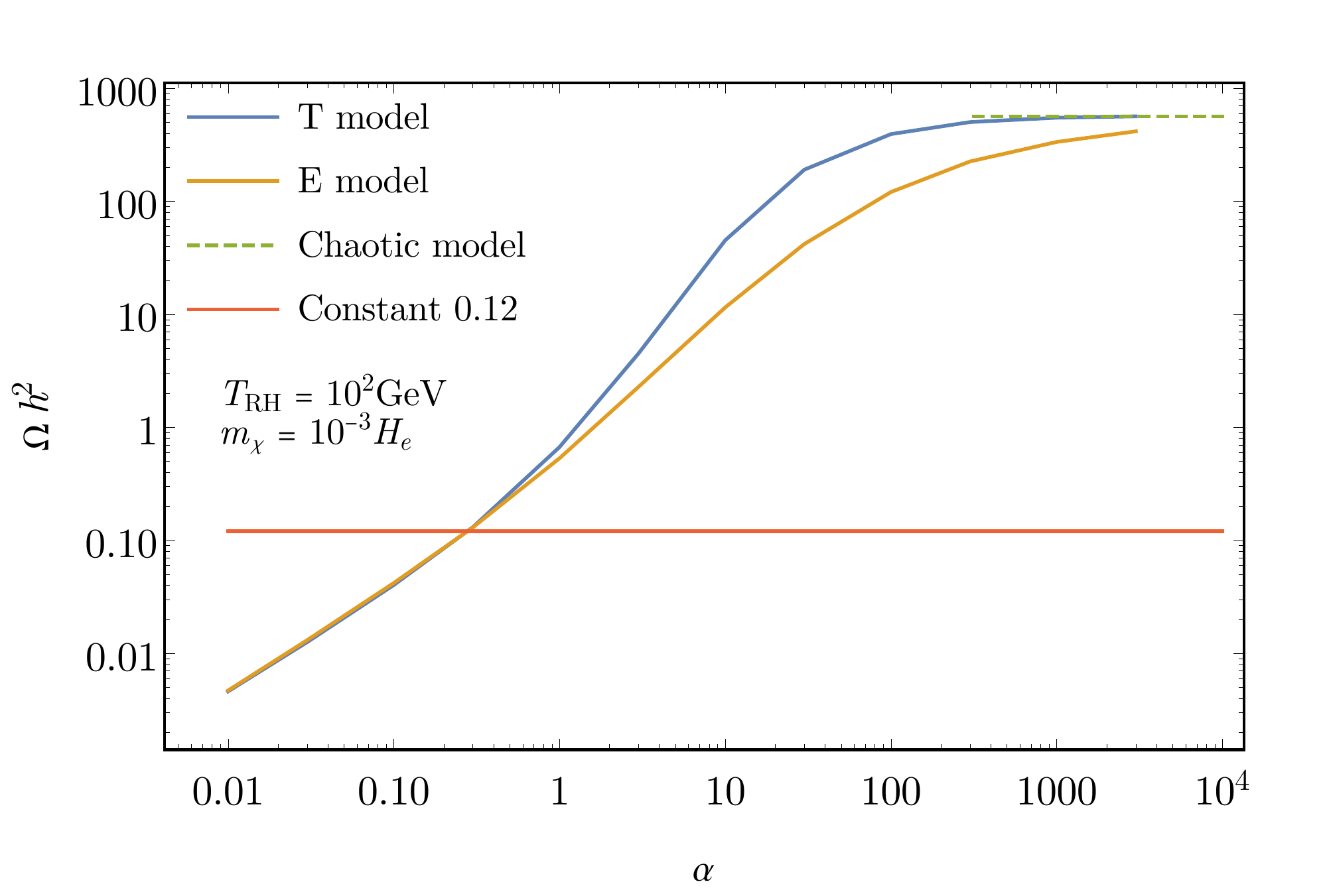} 
\caption{\label{fig:abundance_2}\small
The relic abundance of light ($m_\chi \ll H_e$) scalar spectator particles that arise from gravitational particle production in $\alpha$-attractor models of inflation.  The relic abundance $\Omega h^2$ is insensitive to $m_\chi$ for $m_\chi \ll H_e$, as seen in \fref{fig:abundance_1}, and we take $m_\chi = 10^{-3} H_e$ here.  We take $T_\RH = 10^2 \GeV$, and for other reheating temperatures $\Omega h^2 \propto T_\RH$ as in \eref{eq:Omegah2}.  The blue and orange curves show the predicted relic abundance $\Omega h^2$ for the T-model and E-model $\alpha$-attractors, respectively.  For comparison we show the measured dark matter relic abundance $\Omega_\mathrm{dm} h^2 \simeq 0.12$ (red line) and the prediction of quadratic chaotic inflation (green-dashed line), which is the limit of the T- and E-model $\alpha$-attractors for $\alpha \to \infty$.  
}
\end{center}
\end{figure}

To illustrate how the efficiency of gravitational particle production varies across $\alpha$-attractor models of inflation, we have calculated $\Omega h^2$ for a light, scalar spectator, and we present these results in \fref{fig:abundance_2}.  
In the regime $m_\chi \ll H_e$, the relic abundance becomes insensitive to $m_\chi$, and $\Omega h^2$ is controlled primarily by $\alpha$ for either the T-model or E-model $\alpha$-attractor.  
Provided that the duration of reheating is sufficiently long \pref{eq:low_TRH}, the dependence on $T_\RH$ is just an overall scaling, $\Omega h^2 \propto T_\RH$ as seen in \eref{eq:Omegah2}.  
We can use \pref{eq:low_TRH} to check that our results are consistent with the late reheating regime:  typical values of \(m_\chi\) give \(8 \times 10^8 \GeV \sqrt{m_\chi / \GeV} \sim 10^{15}\GeV\).

\subsection{Isocurvature}\label{sub:isocurvature}

\begin{figure}[t]
\begin{center}
\includegraphics[width=\textwidth]{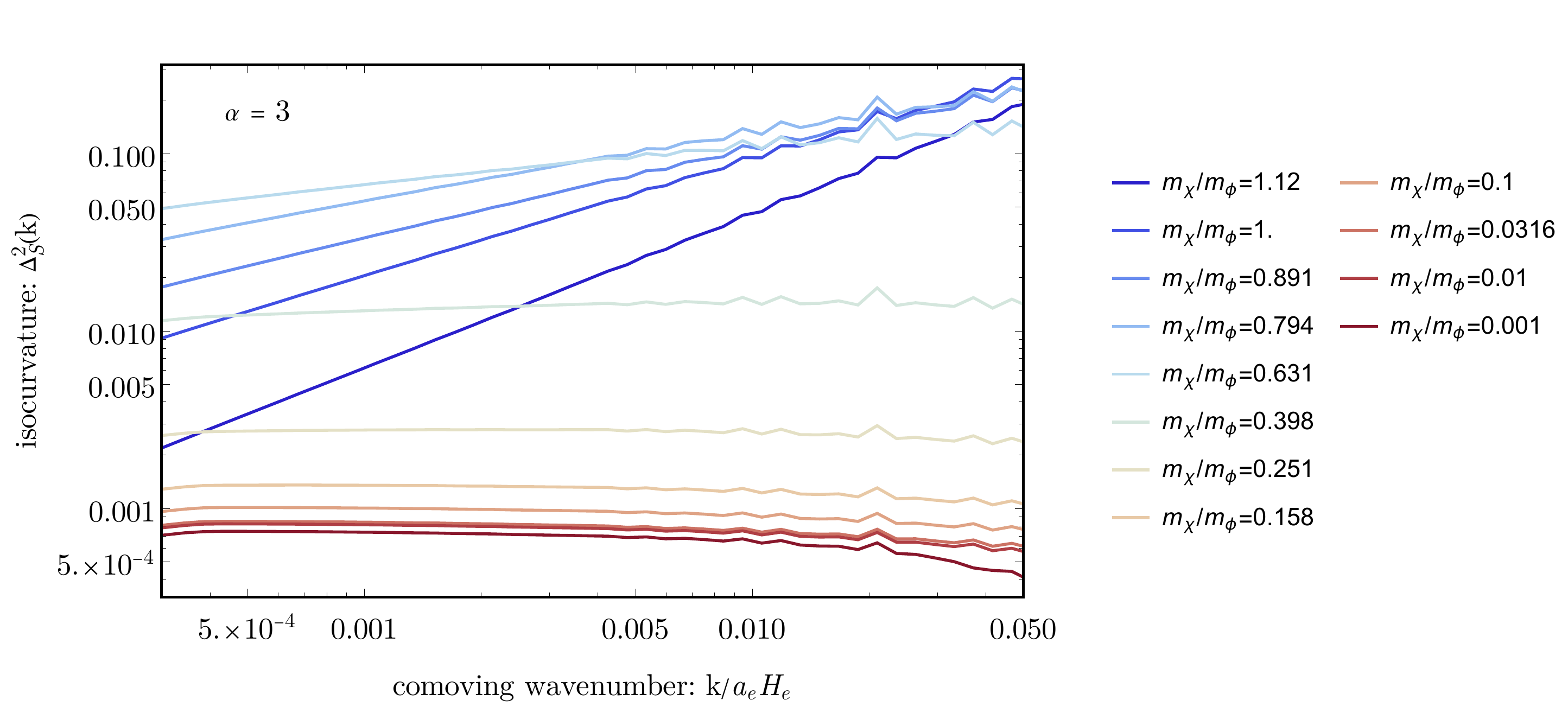}
\caption{\label{fig:isocurvature_1}\small
The predicted isocurvature power spectrum $\Delta_\Scal^2(k)$.  As the spectator mass is raised from $m_\chi / m_\phi \ll 1$ to $m_\chi / m_\phi \gtrsim 1$, the spectrum's low-$k$ tail changes from a red spectral tilt (red-colored curves) to a blue spectral tilt (blue-colored curves).  For this example we take the T-model $\alpha$-attractor with $m_\phi \approx 6 \times 10^{-6} M_p$ and $\alpha = 3$.
}
\end{center}
\end{figure}

We calculate the isocurvature power spectrum by performing the wavenumber and angular integrals in \eref{eq:isocurvature}.  
The mode functions were obtained numerically, and we run into numerical problems when $k/a_e H_e$ is much smaller than $O(1)$.  
This is because the low-$k$ modes leave the horizon earlier, and the mode functions must be solved over a longer time interval, corresponding to many more oscillations of the mode functions, in order to get accurate numerical results.
To avoid these issues, we restrict our attention to comoving wavenumbers in the range $10^{-4} \lesssim k / a_e H_e \lesssim 1$, which corresponds to the modes that left the horizon within $10$ e-foldings before the end of inflation.  
An illustrative selection of our results appears in \fref{fig:isocurvature_1}, where we show the predicted dark matter-photon isocurvature power spectrum for an $\alpha$-attractor model with $\alpha = 3$.  
If the spectator field is light ($m_\chi \ll m_\phi$), then the power spectrum is slightly red-tilted at low $k$, associated with the scale-invariance violation of the slowly-rolling inflaton field.  
As $m_\chi / m_\phi$ becomes larger than $O(1)$, the power spectrum develops a blue tilt~\cite{Chung:2004nh,Chung:2011xd}.  
Additionally the amplitude of the power spectrum is suppressed, which is the same suppression that we encountered for the spectrum in \sref{sub:spectrum}.  

Measurements of the cosmic microwave background have provided strong constraints on the dark matter-photon isocurvature for the modes that are observed in the CMB.  
Currently the strongest constraints come from \textit{Planck},\footnote{Here we have used $\beta_\mathrm{iso}(k_\mathrm{low}) < 0.035$ at 95\%CL (``axion I'' \textit{Planck} TT, TE, EE+lowE+lensing)~\cite{Akrami:2018odb} where $\beta_\mathrm{iso} = \Delta_\Scal^2 / (\Delta_\zeta^2 + \Delta_\Scal^2)$ and $\Delta_\zeta^2 = A_s \simeq 2.1 \times 10^{-9}$.  \textit{Planck} provides limits on several different models of axion-type isocurvature.  The ``axion I'' model was chosen since it implies an isocurvature spectral index of \(n_{\mathcal{I}\mathcal{I}} - 1 \approx 0\), and this is comparable to the isocurvature spectral index that we obtained numerically, \(0.4\).} which correspond to $\Delta_\Scal^2(k_\cmb) < \Delta_\Scal^2(k_\cmb)^\mathrm{max} = 7.3 \times 10^{-11}$ at a pivot scale of $k_\cmb = 0.002 \Mpc^{-1} a_0$.  
The pivot scale  corresponds to roughly $k_\cmb / a_e H_e \sim e^{-50} \simeq 2 \times 10^{-22}$ where we've used \eref{eq:k_cmb}.  
To compare our numerically-evaluated isocurvature at $10^{-4} \lesssim k / a_e H_e$ with these constraints we perform a power-law extrapolation to low-$k$.  
This extrapolation is expected to be very reliable, since the Hubble parameter only changes by a factor of $1 \lesssim H_{50} / H_{10} < 2.1$ across this range of modes.  

\subsection{Constraints}\label{sub:constraints}

\begin{figure}[t!]
\begin{center}
\includegraphics[width=0.8\textwidth]{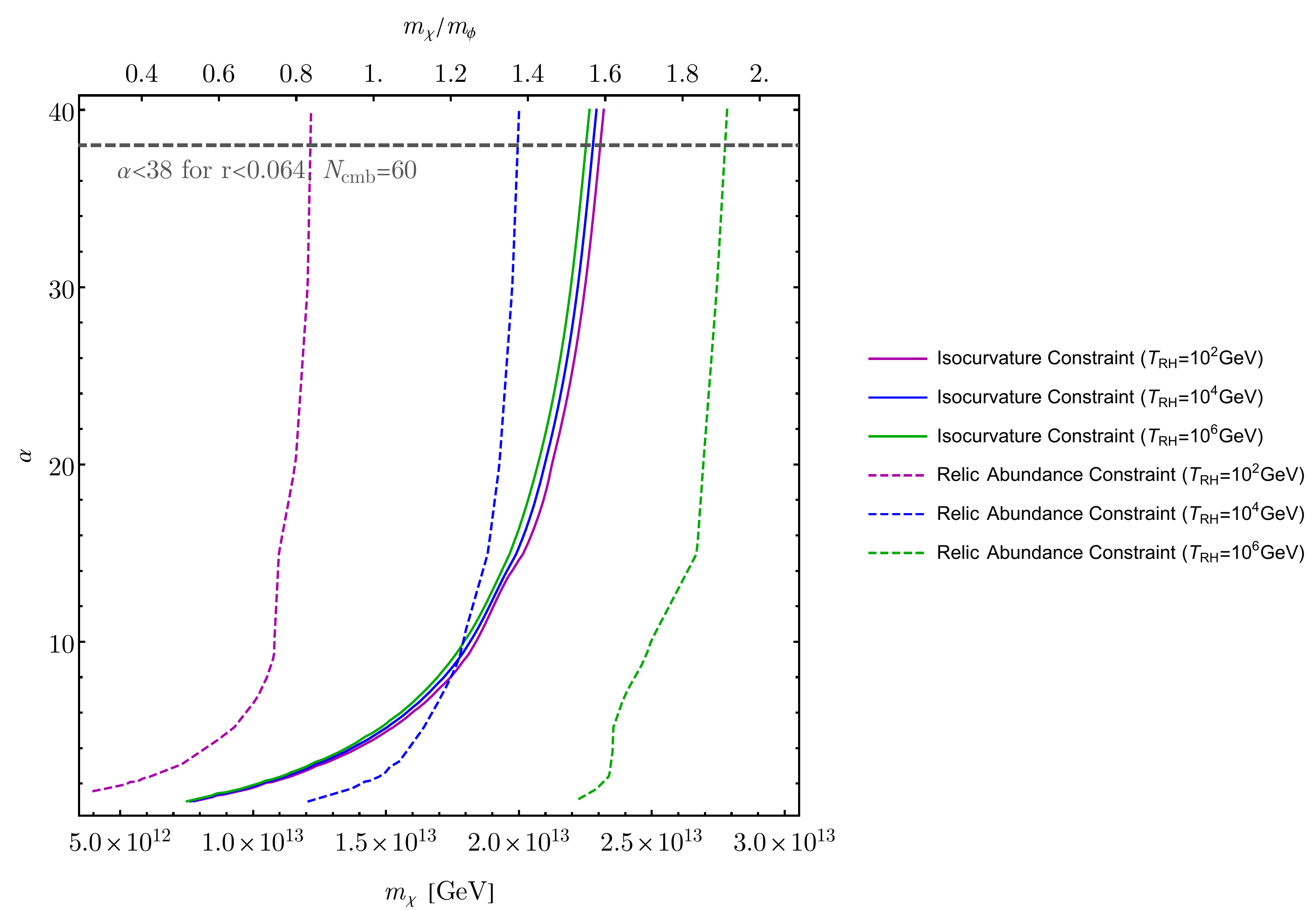} 
\caption{\label{fig:isocurvature_2}\small
A slice of the parameter space with various constraints.  We show the scalar spectator field's mass $m_\chi$, the T-model $\alpha$-attractor's parameter $\alpha$, and we fix $m_\phi = 6 \times 10^{-6} \, M_p$.  The solid curves denote the minimum \(m_\chi\) and maximum \(\alpha\) at a given reheating temperature \(T_\RH\) for which the predicted isocurvature is consistent with the upper bound on \(\Delta_\Scal^2(k_\cmb)\) inferred by \textit{Planck}.  Along the dashed curves, the scalar spectator's predicted relic abundance saturates the measured dark matter relic abundance \(\Omega h^2 \simeq 0.12\), and the spectator is overproduced for smaller $m_\chi$ or larger $\alpha$ at a given $T_\RH$.  The gray-dashed line denotes the maximum \(\alpha\) for which \textit{Planck} and \textit{BICEP/Keck}'s constraint on the tensor-to-scalar ratio \(r < 0.064\) (95\% CL) is satisfied at \(N_\cmb = 60\), and the limit strengthens to \(\alpha \lesssim 20\) at \(N_\cmb = 50\).  Note that the parameter space shown here is consistent with the limit in \eref{eq:low_TRH}.  
}
\end{center}
\end{figure}

We summarize our constraints on this model in \fref{fig:isocurvature_2}.  
Here, the dashed curves give the points in the \(\{m_\chi, \alpha\}\) parameter space where the relic abundance constraint \(\Omega h^2 = 0.12\) is satisfied exactly; namely, the dashed curves show the cases in which \(\chi\) makes up all the dark matter.
Since gravitational particle production \(a^3 n / a_e^3 H_e^3\) decreases as \(m_\chi\) is increased, every point on the left of the dashed curves are excluded, and for each \(\alpha\) we obtain a lower limit on \(m_\chi\).
From the figure we can conclude that the constraint on \(m_\chi\) is the strongest as \(\alpha \to \infty\), and it grows weaker as \(\alpha\) is lowered.
For \(\alpha=10\) and \(T_\RH = 10^4 \GeV\), we get a limit of \(m_\chi \gtrsim 1.2 m_\phi \approx 1.8\times 10^{13} \GeV\).

The solid curves in \fref{fig:isocurvature_2} show the surfaces of maximal CMB-scale isocurvature, $\Delta_\Scal^2(k_\cmb) = 7.3 \times 10^{-11}$.
As explained in \sref{sub:isocurvature}, at a fixed \(\alpha\) the isocurvature amplitude \(\Delta_\Scal^2(k_\cmb)\) decreases for increasing \(m_\chi\), so, similar to the case for relic abundance constraints, we have a lower limit on \(m_\chi\) at each \(\alpha\).
Note that the mass constraints due to isocurvature have a slight dependence on \(T_\RH\): the mass constraints are weaker when \(T_\RH\) is higher.
This shift is due to the \(k_\cmb\) dependence on \(T_\RH\).
If \(T_\RH\) is increased, we know from \eref{eq:k_cmb} that \(k_\cmb\) must decrease accordingly, and the isocurvature \(\Delta_\Scal^2(k_\cmb)\) must also decrease since the spectrum is blue-tilted.
The mass \(m_\chi\) for which the isocurvature constraint is exactly satisfied must then be lowered.
For \(\alpha=20\) and \(T_\RH = 10^6 \GeV\), we get a limit of \(m_\chi \gtrsim 1.4 m_\phi \approx 2.1\times 10^{13} \GeV\).  

Assuming \(T_\RH = 10^9 \GeV\), we also give two empirical fitting formulas for the mass constraint due to isocurvature as functions of \(\alpha\):
\begin{subequations}
\begin{align}\label{eq:empirical_formula_iso}
	m_\chi &\geq m_\infty - \frac{1.1 \times 10^{14} \GeV}{\alpha + 5.4}\\
	\frac{m_\chi}{H_e} &\geq \frac{m_\infty}{H_\infty} - \frac{14.3}{\alpha + 5.7}
\end{align}
\end{subequations}
where \(m_\infty= 2.5 \times 10^{13} \GeV\) and \(H_\infty= 7.3 \times 10^{12} \GeV\) are the mass constraint and the Hubble scale at the end of inflation for chaotic inflation,
corresponding to \(\alpha\to\infty\).

To cross-check our results against previous studies, we consider the limit \(\alpha \to \infty\), which corresponds to chaotic inflation with \(V = m_\phi^2 \phi^2 / 2\) and \(m_\phi = 1.7 \times 10^{13} \GeV\).  
For \(T_\RH = 10^4 \GeV\) we find that the isocurvature constraint imposes \(m_\chi > 3.4H_e \approx 1.7 m_\phi\).  
This calculation agrees well with an earlier studied of isocurvature that obtains \(m_\chi \gtrsim 2.1 m_\phi\) for the same model~\cite{Chung:2011xd} (see also \rref{Chung:2004nh}).  

\section{Conclusion}\label{sec:Conclusion}

In this article we have studied the gravitational production of superheavy dark matter in an inflationary cosmology.  
In particular, we have focused on spin-0 dark matter and the $\alpha$-attractor models of inflation, considering both the T-model and E-model $\alpha$-attractors.  
In this background we construct the mode equations for a scalar spectator field $\chi$ that is minimally-coupled to gravity, and we solve these mode equations assuming a Bunch-Davies initial condition to determine the scalar field's mode functions.  
From these mode functions we extract the Bogoliubov coefficients, which are then used to calculate the relic abundance and isocurvature power spectrum of the $\chi$ particles.  

The key results of this study are summarized as follows.  
In the regime $\alpha \gg 1$, we reproduce the known results for quadratic chaotic inflation, which is a check of our numerical approach.  
Lowering $\alpha$ causes the energy scale of inflation to become smaller, and our calculation shows that there is a corresponding reduction in gravitational particle production when \(T_\mathrm{RH}\) is held fixed; see \fref{fig:abundance_2}.
Alternatively a smaller \(\alpha\) can be offset by an increased \(T_\mathrm{RH}\) to leave the relic abundance unchanged.
This result agrees with the expected scaling behavior based on dimensional analysis.  
We have also calculated the isocurvature power spectrum, which appears in \fref{fig:isocurvature_1}.  
For a light spectator, $m_\chi \ll m_\phi$, we observe a slightly red-tilted power spectrum, which turns over into a blue-tilted spectrum as $m_\chi$ is raised above the inflaton mass $m_\phi$.  
To avoid observational constraints on dark matter isocurvature, with the strongest limits provided by \textit{Planck}, the power spectrum must be sufficiently blue-tilted.
This translates into a lower limit on the spectator's mass; see \fref{fig:isocurvature_2} and the empirical fitting formulas in \eref{eq:empirical_formula_iso}.  

The work presented in this article demonstrates the continued viability of WIMPzilla dark matter, particularly a superheavy and minimally-coupled scalar.   
Many previous numerical studies of WIMPzilla dark matter have focused on quadratic chaotic inflation, which is now ruled out by \textit{Planck}.  
By extending these studies to the $\alpha$-attractor class of models, we have framed WIMPzilla dark matter in the context of a compelling and testable inflationary cosmology, namely $\alpha$-attractor inflation.  

\subsubsection*{Acknowledgements}
We are grateful to Mustafa Amin, Mudit Jain, Edward Kolb, Evan McDonough, Zong-Gang Mou, Evangelos Sfakianakis, and Hong-Yi Zhang for discussions and comments on the draft.  We especially thank Daniel Chung for illuminating discussions of isocurvature.

\bibliographystyle{JHEP}
\bibliography{alpha_attractor}

\end{document}